\documentclass[fleqn,usenatbib]{mnras}

\usepackage{newtxtext,newtxmath}

\usepackage[T1]{fontenc}
\usepackage{ae,aecompl}

\usepackage{graphicx}	
\usepackage{amsmath}	
\usepackage{amssymb}	

\usepackage{bm}

\title[The Crab Nebula with tangled magnetic field]{Confinement of the Crab Nebula with tangled magnetic field by its supernova remnant}

\author[S. J. Tanaka, K. Toma, \& N. Tominaga]{
Shuta J. Tanaka,$^{1,2}$\thanks{E-mail: sjtanaka@center.konan-u.ac.jp (SJT)}
Kenji Toma,$^{3,4}$
and Nozomu Tominaga$^{1,5}$
\\
$^{1}$Department of Physics, Faculty of Science and Engineering, Konan University, 8-9-1 Okamoto, Kobe, Hyogo 658-8501, Japan\\
$^{2}$Department of Physics and Mathematics, Aoyama Gakuin University, 5-10-1 Fuchinobe, Sagamihara, Kanagawa 252-5258, Japan\\
$^{3}$Frontier Research Institute for Interdisciplinary Sciences, Tohoku University, Sendai 980-8578, Japan\\
$^{4}$Astronomical Institute, Tohoku University, Sendai 980-8578, Japan\\
$^{5}$Kavli Institute for the Physics and Mathematics of the Universe, The University of Tokyo, 5-1-5 Kashiwanoha, Kashiwa, Chiba 277-8583, Japan
}

\date{Accepted XXX. Received YYY; in original form ZZZ}

\pubyear{2018}

\begin{document}
\label{firstpage}
\pagerange{\pageref{firstpage}--\pageref{lastpage}}
\maketitle

\begin{abstract}
	A pulsar wind is a relativistic outflow dominated by Poynting energy at its base.
	Based on the standard ideal magnetohydrodynamic (MHD) model of pulsar wind nebulae (PWNe) with the ordered magnetic field, the observed slow expansion $v_{\rm PWN} \ll c$ requires the wind to be dominated by kinetic energy at the upstream of its termination shock, which conflicts with the pulsar wind theory ($\sigma$-problem).
	In this paper, we extend the standard model of PWNe by phenomenologically taking into account conversion of the ordered to turbulent magnetic field and dissipation of the turbulent magnetic field.
	Disordering of the magnetic structure is inferred from the recent three-dimensional relativistic ideal MHD simulations, while magnetic dissipation is a non-ideal MHD effect requiring a finite resistivity.
    We apply this model to the Crab Nebula and find that the conversion effect is important for the flow deceleration, while the dissipation effect is not.
    Even for Poynting-dominated pulsar wind, we obtain the Crab Nebula's $v_{\rm PWN}$ by adopting a finite conversion time-scale of $\sim 0.3$ yr.
	Magnetic dissipation primarily affects the synchrotron radiation properties.
	Any values of the pulsar wind magnetization $\sigma_{\rm w}$ are allowed within the present model of the PWN dynamics alone, and even a small termination shock radius of $\ll 0.1$ pc reproduces the observed dynamical features of the Crab Nebula.
	In order to establish a high-$\sigma_{\rm w}$ model of PWNe, it is important to extend the present model by taking into account the broadband spectrum and its spacial profiles.
\end{abstract}

\begin{keywords}
	magnetic fields ---
	MHD ---
	radiation mechanisms: non-thermal ---
	pulsars: individual (Crab pulsar) ---
	ISM: individual objects (Crab Nebula) ---
	turbulence ---
\end{keywords}



\section{Introduction}\label{sec:intro}

Extraction of rotational energy by magnetic braking is the most promising mechanism driving relativistic outflows of active galactic nuclei, gamma-ray bursts and pulsar winds.
These outflows must be Poynting dominated, i.e., the magnetization parameter $\sigma$ (the ratio of Poynting to particle energy fluxes) is much larger than unity at their base.
In the theoretical point of view, it is difficult to reduce $\sigma$ with accelerating the pulsar wind within the ideal magnetohydrodynamic (MHD) formulation \citep[c.f.,][]{Komissarov11, Lyubarsky11}.
However, the observations of the pulsar wind nebulae (PWNe) show two important properties which suggest that the pulsar winds are low-$\sigma$ outflows at their terminal, i.e., at the pulsar wind termination shock \citep{Rees&Gunn74, Kennel&Coroniti84a, Atoyan&Aharonian96}.
One is the broadband spectrum argument which requires $\sigma \ll 1$ in order to reproduce the observed high radiative efficiency and ratio of the X-ray to $\gamma$-ray fluxes, that is spectral $\sigma$-problem.
The other is the nebular dynamics argument which requires $\sigma \ll 1$ in order to reproduce the observed expansion velocity of the Crab Nebula $v_{\rm PWN} \ll c$, that is dynamical $\sigma$-problem.
In this paper, we focus on the dynamical $\sigma$-problem.

A possible solution is dissipation of the magnetic field by magnetic reconnection before the pulsar wind reaches to the termination shock.
Because, in general, pulsars are an oblique rotator, i.e., the magnetic axis is inclined with respect to the rotation axis, there are alternating current sheets around the equatorial region, called striped wind \citep{Coroniti90, Bogovalov99}.
Although the alternating magnetic field configuration seems an ideal situation for magnetic reconnection, the dissipation time-scale increases with accelerating the flow by the relativistic time dilation.
\citet{Lyubarsky&Kirk01} concluded that the flow does not become low-$\sigma$ before it reaches the termination shock for the case of the Crab \citep[c.f.,][]{Kirk&Skjaeraasen03}.
\citet{Lyubarsky03b} proposed driven magnetic reconnection at the termination shock and the numerical studies showed that this mechanism is a viable \citep{Sironi&Spitkovsky11, Amano&Kirk13}.
However, we require the Crab pulsar to be an almost orthogonal rotator in order to obtain the Crab Nebula's value $\sigma \sim 10^{-3}$, i.e., in order to annihilate almost all the magnetic field \citep{Komissarov13}.
This is inconsistent with the pulse profiles of the high energy emission from the Crab pulsar \citep[e.g.,][]{Harding+08}.

The above difficulties to obtain $\sigma \ll 1$ at the upstream of the termination shock remind us a possibility of magnetic dissipation after the termination shock, i.e., magnetic dissipation inside the Crab Nebula \citep{Komissarov13}.
It means that we have to abandon the familiar conclusion which $\sigma \sim 10^{-3}$ at the upstream of the termination shock and look for a solution of the dynamical $\sigma$-problem with setting $\sigma \gtrsim 1$ at the termination shock.
Therefore, we should extend the standard MHD model of the PWN developed by \citet{Kennel&Coroniti84a, Kennel&Coroniti84b} (hereafter, KC model).

So far, the high-$\sigma$ model of PWNe are studied by \citet{Lyutikov10} and \citet{Zrake&Arons17}.
\citet{Lyutikov10} adopted a steady-state, axially symmetric, `non-resistive' force-free model of a PWN, but set magnetic dissipation regions at the polar and equatorial boundaries in their model.
They obtained a solution which a high-$\sigma$ wind from the pulsar can be confined by a non-expanding spherical cavity, although some physics are missed in their simple model, for example, the magnetic dissipation regions are not the plasma heating region, but are just an energy sink boundary.
\citet{Zrake&Arons17} considered a steady-state, spherically symmetric, non-ideal MHD model of a PWN, i.e., they extended KC model including magnetic dissipation in a phenomenological way adopted from \citet{Drenkhahn02}.
They also found solutions which reproduce the observed expansion velocity of the Crab Nebula even when the magnetization of the pulsar wind is much larger than unity.
The ordered toroidal magnetic field directly dissipates into the heat of plasma in their formulation.
As they consider a coalescence instability of turbulent magnetic structures (turbulent relaxation model), they implicitly assume the existence of the turbulent magnetic field in addition to the ordered one.

There are also some other motivations to introduce the turbulent magnetic field.
One is the observed polarization of the Crab Nebula \citep[e.g.,][]{Aumont+10, Moran+13, Chauvin+17}.
The synchrotron emission has the polarization fraction of $\sim 20 - 30 \%$ indicating that the magnetic field is not perfectly ordered.
In order to lower the polarization fraction well below $\sim 70 \%$ \citep[e.g.,][]{Rybicki&Lightman79}, the ratio of ordered to disordered magnetic field should be order unity \citep[c.f.,][]{Bucciantini+05, Bucciantini+17}.
The other motivation is the torus structure observed in X-ray \citep[][]{Weisskopf+00}.
\citet{Shibata+03} discussed that the X-ray surface brightness map based on KC model becomes `lip-shaped' rather than torus.
They reproduced the observed torus morphology by assuming that the turbulent magnetic field is comparable to the toroidal one.
We expect the coexistence of the toroidal and turbulent magnetic field inside PWNe.
Note that the magnetic dissipation model by \citet{Zrake&Arons17} does not account for such polarization and surface brightness properties because their model has no turbulent magnetic field everywhere in the nebula, while we introduce a finite strength of the turbulent magnetic field as an intermediate state before the ordered magnetic field dissipating into the heat of plasma.

We construct a basic picture of a magnetic field configuration inside PWNe by observing the three-dimensional ideal MHD simulation by \citet{Porth+14a} as follows.
The pulsar wind has the pure toroidal magnetic field and keeps injecting it into the inner boundary (termination shock) of the PWN.
Such a magnetic field configuration becomes turbulent at least several shock radii flowing away from the termination shock by the magnetic kink instability \citep[c.f.,][]{Begelman98, Mizuno+11}.
Outer PWN also suffers from the turbulence induced by the Rayleigh-Taylor instability between the PWN and supernova ejecta \citep[e.g.,][]{Hester+96, Porth+14b}.
Ordered toroidal magnetic field at inner PWNe is gradually converted into the turbulent one along with the nebular flow, and subsequently the turbulent one dissipates by magnetic reconnection \citep[e.g,][]{Lazarian&Vishniac99, Takamoto&Lazarian16}.
In contrast to the previous studies, we consider that magnetic dissipation proceeds gradually rather than instantaneously after the conversion from the ordered magnetic field.
The turbulent magnetic field exists as an intermediate state before magnetic dissipation.
Our formulation is useful in order to understand the observations and multi-dimensional numerical studies of PWNe.

In this paper, we separately study the effects of the turbulent magnetic field and its dissipation inside a PWN.
In section \ref{sec:Model}, we describe our model of a nebular flow.
A heuristic formulation is adopted by reference to the past studies, and the mathematical details are found in Appendices in the covariant form.
In section \ref{sec:Application}, we apply the model to the Crab Nebula by requiring the velocity of the nebular flow at the outer boundary to be the observed value of $v_{\rm PWN} = 1500~{\rm km~s^{-1}}$.
The results are discussed in section \ref{sec:Discussion} and we conclude in section \ref{sec:Conclusions}.

\section{Model}\label{sec:Model}

Here, we describe our model of a PWN outflow.
We extend KC model by including the turbulent component of the magnetic field.
Conversion of the toroidal magnetic field to the turbulent one and dissipation of the turbulent magnetic field are modeled phenomenologically.

\subsection{Inner boundary of nebular flow}\label{sec:InnerBoundaryOfNebulaFlow}

We consider that the nebular flow is a shocked pulsar wind and we adopt the upstream properties from KC model.
The pulsar wind of KC model is a `laminar' relativistic outflow composed of a cold $e^{\pm}$ and the `pure toroidal' magnetic field at the termination shock.
The power of the pulsar wind $L_{\rm wind}$ is characterized by three non-dimensional parameters as
\begin{eqnarray}\label{eq:WindLuminosity}
	L_{\rm wind}
	& = &
	\kappa_{\rm w} \dot{N}_{\rm GJ} \gamma_{\rm w} m_{\rm e} c^2 (1 + \sigma_{\rm w}),
\end{eqnarray}
where $\kappa_{\rm w}$ is the wind pair multiplicity (the particle number flux normalized by the Goldreich-Julian number flux $\dot{N}_{\rm GJ} \equiv \sqrt{6 c L_{\rm spin}}/e$), $\gamma_{\rm w}$ is the wind bulk Lorentz factor, $\sigma_{\rm w}$ is the wind magnetization parameters (the ratio of the Poynting to the kinetic energy fluxes) and $L_{\rm spin}$ is the spin-down luminosity of the pulsar.
Setting $L_{\rm wind} = L_{\rm spin}$, we obtain one constraint,
\begin{eqnarray}\label{eq:WindProperties}
	\kappa_{\rm w} \gamma_{\rm w} (1 + \sigma_{\rm w})
	& = &
	1.4 \times 10^{10}
	\left( \frac{L_{\rm spin}}{10^{38}~{\rm erg~s^{-1}}} \right)^{\frac{1}{2}}.
\end{eqnarray}
For a given $L_{\rm spin}$, the two of three parameters, $\kappa_{\rm w}$, $\gamma_{\rm w}$, and $\sigma_{\rm w}$, represent the pulsar wind property completely \citep[c.f.,][]{Tanaka&Takahara13a}.

We assume that the termination shock of the pulsar wind is a strong shock at $r = r_{\rm TS}$.
One of the important assumptions of KC model is that the shock is a standing shock ($r_{\rm TS} =$ const.), i.e., the inner edge of the nebular flow is at rest in the observer (pulsar) frame.
Using the Rankin-Hugoniot relations summarized in Appendix \ref{app:JumpCondition}, the inner boundary condition for the nebular flow is determined from the pre-shock property of the pulsar wind satisfying equation (\ref{eq:WindProperties}).

\subsection{Basic equations}\label{sec:BasicEquations}

The nebular flow of KC model has the following five assumptions: (i) steady state, (ii) spherical symmetry, (iii) pure radial flow, (iv) pure toroidal magnetic field, and (v) ideal MHD.
Their nebular flow has five variables, the proper enthalpy density $w$, the pressure $p$, the proper density $n$, the radial (four) velocity $u$, and the (ordered) toroidal magnetic field $\bar{B} = \gamma \bar{b}$ in the observer frame.
As is shown in Appendix \ref{app:KC84}, adopting the equation of state equation (\ref{eq:app:EOS}), the system is described by the four equations (equations (\ref{eq:app:NumberConservationKC84}) -- (\ref{eq:app:InductionEquationKC84})).
All of the four equations are integrable and the nebular flow is described by one algebraic equation \citep[equation (5.7) of][]{Kennel&Coroniti84a}.

In this paper, we omit the two of the above five assumptions (iv) and (v) of KC model, i.e., we consider the turbulent magnetic field and its dissipation.
We adopt phenomenological descriptions of these effects referring to \citet{Drenkhahn02} and \citet{Drenkhahn&Spruit02}.
Instead of the assumption (iv), we assume that the initially pure toroidal magnetic field is being converted to the turbulent magnetic field with the rate $\tau^{-1}_{\rm conv}$ along with the nebular flow.
We also assume, instead of the assumption (v), that only the turbulent component dissipates to the heat of plasma with the rate $\tau^{-1}_{\rm diss}$.
In addition, we introduce the radiative cooling $\Lambda_{\rm rad}$ in order to study the basic properties of radiation, such as the radial brightness profile.
In section \ref{sec:RadiativeProperties}, we will find that the radiative cooling hardly changes the flow dynamics for typical parameters of the Crab Nebula.
Mathematical details of our formulation are described in Appendix \ref{app:Derivation}.

We introduce the sixth variable, the strength of the turbulent magnetic field $\delta b^2$, in addition to the five variables of KC model, $w, p, n, u$, and $\bar{b}$.
The turbulent magnetic field $\delta {\bm b}$ is assumed to be isotropic in the flow proper frame satisfying $\langle \delta {\bm b} \rangle = 0$ and $\langle \delta {\bm b}^2 \rangle = \delta b^2$, where $\langle \rangle$ represents ensemble average (see Appendix \ref{app:TurbulentMagneticField}).
The corresponding six equations are two algebraic equations;
\begin{eqnarray}
	w
	& = &
	n m_{\rm e} c^2 + \frac{\hat{\Gamma}}{\hat{\Gamma} - 1} p
	, \label{eq:hotEOS} \\
	4 \pi n u c r^2
	& = &
	\kappa_{\rm w} \dot{N}_{\rm GJ}
	=
	{\rm const.}
	, \label{eq:NumCons}
\end{eqnarray}
and four differential equations;
\begin{eqnarray}
	\frac{d}{d r} \left[r^2 \gamma u \left(w + \bar{b}^2 + \frac{2}{3} \delta b^2 \right) \right]
	& = &
	- r^2 \gamma \frac{\Lambda_{\rm rad}}{c}
	, \label{eq:TotalEnergyConservation} \\
	\frac{d}{d r} (w - p) + w \frac{d}{d r} \ln u r^2
	& = &
	\frac{\delta b^2 / 2}{u c \tau_{\rm diss}}
	- \frac{\Lambda_{\rm syn}}{u c}
	, \label{eq:InternalEnergyConservation} \\
	\frac{d}{d r} \frac{\bar{b}^2}{2} + \frac{\bar{b}^2}{2} \frac{d}{d r} \ln u^2 r^2
	& = &
	- \frac{\bar{b}^2 / 2}{u c \tau_{\rm conv}}
	, \label{eq:ToroidalMagneticEnergyConservation} \\
	\frac{d}{d r} \frac{\delta b^2}{2} + \frac{2}{3} \delta b^2 \frac{d}{d r} \ln u r^2
	& = &
	- \frac{\delta b^2 / 2}{u c \tau_{\rm diss}}
	+ \frac{\bar{b}^2  / 2}{u c \tau_{\rm conv}}
	, \label{eq:TurbulentMagneticEnergyConservation}
\end{eqnarray}
where the magnetic field is divided by $\sqrt{4 \pi}$ and the velocity is normalized by the speed of light $c$ for notational convenience.
When we omit the left-hand side terms of equation (\ref{eq:TurbulentMagneticEnergyConservation}), we recover \citet{Zrake&Arons17} where $\bar{b}$ directly dissipates to the heat of plasma (precisely, $\delta b^2 = 0$ in equation (\ref{eq:TotalEnergyConservation}), see section \ref{sec:SpectrumOfTurbulence}).
KC model corresponds to $\delta b^2 = 0$ without all the right-hand side terms of equations (\ref{eq:TotalEnergyConservation}) -- (\ref{eq:ToroidalMagneticEnergyConservation}).
In this paper, we set the adiabatic index of $\hat{\Gamma} = 4 / 3$ for a relativistically hot plasma.
The remaining two parameters for the equations, $\tau_{\rm conv}$ and $\tau_{\rm diss}$, are described in section \ref{sec:ParametersOfBasicEqs}.

We only consider the synchrotron cooling for the radiation cooling rate $\Lambda_{\rm rad} \equiv n P_{\rm syn}$, where $P_{\rm syn}$ is the synchrotron cooling rate per particle.
We ignore the other sub-dominant emission mechanisms, such as inverse Compton scattering, for simplicity.
Furthermore, we assume the synchrotron radiation from the mono-energetic particles whose Lorentz factor is $\gamma_{\rm th} \equiv (w - p) / (n m_{\rm e} c^2)$ (see also the discussion in section \ref{sec:BroadbandRadiation}), i.e.,
\begin{eqnarray}\label{eq:SynCoolingRate}
	P_{\rm syn}
	=
	\frac{4}{3} \sigma_{\rm T} c \gamma^2_{\rm th} \frac{\bar{b}^2 + \delta b^2}{2},
\end{eqnarray}
where $\sigma_{\rm T}$ is the Thomson cross section.

\subsection{Characteristics of turbulent magnetic field}\label{sec:CharacteristicsOfTurbulentMagneticField}

Introduction of the turbulent magnetic field is a unique feature of our formulation.
On the present description of the turbulent magnetic field, the turbulent magnetic field behaves as a relativistically hot gas, i.e., a photon gas.
Neglecting the right-hand sides of equations (\ref{eq:InternalEnergyConservation}) and (\ref{eq:TurbulentMagneticEnergyConservation}), we obtain
\begin{eqnarray}
	\frac{d}{d r} (w - p) - \frac{w}{n} \frac{d n}{d r}
	& = &
	0
	, \label{eq:SimplifiedInternalEnergyConservation} \\
	\frac{d}{d r} \frac{\delta b^2}{2} - \frac{2}{3} \frac{\delta b^2}{n} \frac{d n}{d r}
	& = &
	0
	, \label{eq:SimplifiedTurbulentMagneticEnergyConservation}
\end{eqnarray}
respectively.
For a relativistically hot plasma of $\hat{\Gamma} = 4/3$, we obtain $p \propto n^{4/3}$ from equation (\ref{eq:SimplifiedInternalEnergyConservation}).
Equation (\ref{eq:SimplifiedTurbulentMagneticEnergyConservation}) leads to $\delta b^2 \propto n^{4/3}$.
Comparing each term of equations (\ref{eq:SimplifiedInternalEnergyConservation}) and (\ref{eq:SimplifiedTurbulentMagneticEnergyConservation}), the turbulent magnetic field is considered as a relativistic gas whose energy density of $e_{\delta} = \delta b^2 / 2$ and pressure of $p_{\delta} = e_{\delta} / 3 = \delta b^2 / 6$.
On the other hand, equation (\ref{eq:ToroidalMagneticEnergyConservation}) without the right-hand side gives the conservation of the toroidal magnetic flux $\bar{b}^2 \propto (r u)^{-2} \propto (n r)^2$, which is very different behavior from the above two components.

Interestingly, equation (\ref{eq:TurbulentMagneticEnergyConservation}) can effectively disappear from the equations with the use of the new variables $w' = w + e_{\delta} + p_{\delta}$ and $p' = p + p_{\delta}$.
Adding equation (\ref{eq:InternalEnergyConservation}) to equation (\ref{eq:TurbulentMagneticEnergyConservation}), we have
\begin{eqnarray}
	\frac{d}{d r} \left[r^2 \gamma u \left(w' + \bar{b}^2 \right) \right]
	& = &
	- r^2 \gamma \frac{\Lambda_{\rm rad}}{c}
	, \label{eq:ModifiedTotalEnergyConservation} \\
	\frac{d}{d r} (w' - p') + w' \frac{d}{d r} \ln u r^2
	& = &
	\frac{\delta b^2 / 2}{u c \tau_{\rm conv}}
	- \frac{\Lambda_{\rm syn}}{u c}
	, \label{eq:ModifiedInternalEnergyConservation}
\end{eqnarray}
instead of equations (\ref{eq:TotalEnergyConservation}) and (\ref{eq:InternalEnergyConservation}), where equation (\ref{eq:ToroidalMagneticEnergyConservation}) does not change.
The equations become essentially the same as the case that the toroidal magnetic field dissipates directly to heat up the plasma characterized by $(w',p')$ with the time-scale of $\tau_{\rm conv}$, and then $\tau_{\rm diss}$ disappears \citep[c.f.,][]{Lyubarsky&Kirk01, Drenkhahn&Spruit02, Zrake&Arons17}.
We will discuss our results in this point of views in section \ref{sec:Comparison}.

Nevertheless, $\tau_{\rm diss}$ still has a role.
In equation (\ref{eq:SynCoolingRate}), the magnetic energy density is the sum of the ordered and turbulent magnetic field.
Conversion of the magnetic topology does not change the synchrotron cooling rate, while magnetic dissipation changes.
If we ignore the radiative cooling $\Lambda_{\rm rad}$, the dynamics of the nebular flow is controlled only by $\tau_{\rm conv}$ (see also section \ref{sec:Comparison}).
$\tau_{\rm diss}$ changes radiative properties (see section \ref{sec:RadiativeProperties}).

\subsection{Properties of basic equations}\label{sec:PropertiesOfBasicEqs}

For convenience, we introduce the following five variables, the total enthalpy density $\epsilon$, the total energy flux $L$, the characteristic velocity $\beta_{\rm c}$, the magnetization $\sigma$, and the randomness of the magnetic field $\chi$, as
\begin{eqnarray}
	\epsilon(r)
	& \equiv &
	w + \bar{b}^2 + (2/3) \delta b^2
	, \label{eq:TotalEnergyDensity} \\
	L(r)
	& \equiv &
	4 \pi r^2 \gamma u c \epsilon(r)
	, \label{eq:TotalEnergyFlux} \\
	\beta^2_{\rm c}(r)
	& \equiv &
	\frac{\hat{\Gamma} p + \bar{b}^2 + (2/9)\delta b^2}{\epsilon}
	, \label{eq:CharacteristicVelocity} \\
	\sigma(r)
	& \equiv &
	\frac{\bar{b}^2 + (2/3)\delta b^2}{w}
	, \label{eq:Magnetization} \\
	\chi(r)
	& \equiv &
	\frac{\delta b^2}{\bar{b}^2}
	, \label{eq:Randomness}
\end{eqnarray}
respectively.
These variables are functions of $r$.

We set $L(r_{\rm TS}) = L_{\rm spin}$ and $L(r)$ decreases with radius because of the radiation loss (see equations (\ref{eq:TotalEnergyConservation})).
The physical meaning of the characteristic velocity $\beta_{\rm c}(r)$ will be found in equation (\ref{eq:DifferentialEquationForVelocity}) below, and $\beta^2_{\rm c} \rightarrow 1/3$ for $\bar{b}^2 \rightarrow 0$ while $\beta^2_{\rm c} \sim 1$ for $\bar{b}^2 \gg w + (2/3) \delta b^2$.
$\sigma(r)$ is close to a counterpart of the upstream magnetization $\sigma_{\rm w}$ (equations (\ref{eq:WindLuminosity}) and (\ref{eq:WindProperties})) and is a relativistic invariant while $\sigma_{\rm w}$ is not.
Definition of the upstream magnetization $\sigma_{\rm w}$ includes the upstream four-velocity $u_{\rm w}$ in the denominator (equation (\ref{eq:app:SigmaWind})) and then it is not defined in the proper frame of $u_{\rm w} = 0$, while $\sigma(r)$ is defined only the proper frame values.
Note that $\sigma_{\rm w} \approx \sigma(r_{\rm TS})$ for $u_{\rm w} \gg 1$ (c.f., equation (\ref{eq:app:RHConditionSigma})).
$\chi(r)$ is the ratio of the turbulent to toroidal magnetic energy densities.
$\chi = 3$ is the characteristic value for depolarizing the synchrotron radiation because only one of the three vector components of $\delta {\bm b}$ contributes to depolarization, i.e., two of the three vector components parallel to the line of sight and to the ordered field $\bar{b}$ do not depolarize the synchrotron radiation \citep[c.f.,][]{Bucciantini+17}.

Combining equations (\ref{eq:hotEOS}) -- (\ref{eq:TurbulentMagneticEnergyConservation}), we find the equation describing evolution of the four-velocity
\begin{eqnarray}\label{eq:DifferentialEquationForVelocity}
	\epsilon (\beta^2 - \beta^2_{\rm c}) \frac{d u}{d r}
	& = &
	\frac{2 u}{r} \left( \hat{\Gamma} p + \frac{2}{9} \delta b^2 \right)
	+
	(\hat{\Gamma} - 1) \frac{\Lambda_{\rm rad}}{c}
	\nonumber \\
	& + &
	\frac{\bar{b}^2}{3 c \tau_{\rm conv}}
	+
	\left(\frac{4}{3} - \hat{\Gamma} \right) \frac{\delta b^2}{2 c \tau_{\rm diss}}.
\end{eqnarray}
The characteristic velocity $\beta_{\rm c}$ corresponds to a phase velocity of a fast magnetosonic wave for the case of the pure toroidal magnetic field \citep[e.g.,][]{Lyutikov&Blandford03}.
All the terms in the right-hand side of equation (\ref{eq:DifferentialEquationForVelocity}) is positive and then the nebular flow always decelerates with $r$ since we consider the post shock flow, i.e., $\beta(r_{\rm TS}) < \beta_{\rm c}$ \citep[c.f.,][]{Zrake&Arons17}.

Not only the sign of the right-hand side of equation (\ref{eq:DifferentialEquationForVelocity}), but also the factor in the left-hand side $\epsilon(\beta^2 - \beta^2_{\rm c})$ is important to determine the velocity profile.
Even for KC model which has only the first term ($\delta b^2 = 0$) in the right-hand side of equation (\ref{eq:DifferentialEquationForVelocity}), the flow has two distinct behaviors.
One is $u \propto r^{-2}$ for a low-$\sigma_{\rm w}$ case and the other is $u =$ const. for a high-$\sigma_{\rm w}$ case with the terminal velocity of $\sigma_{\rm w} / \sqrt{1 + 2 \sigma_{\rm w}}$ for both cases \citep[][]{Kennel&Coroniti84a}.
The flow behaves differently from KC model when any one of the last three terms of the right-hand side of equation (\ref{eq:DifferentialEquationForVelocity}) is larger than the first term.

\subsection{Parameters of basic equations}\label{sec:ParametersOfBasicEqs}

We have two parameters of the equations, $\tau_{\rm conv}$ and $\tau_{\rm diss}$.
$\tau_{\rm conv}$ is the most important parameter to change the dynamics of the nebular flow from KC model as already mentioned in section \ref{sec:CharacteristicsOfTurbulentMagneticField}.
In this paper, we assume that $\tau_{\rm conv}$ is a constant in the proper frame.
There would be some other choices of $\tau_{\rm conv}$ and we will discuss in section \ref{sec:Comparison}.

$\tau_{\rm diss}$ does not directly affect the dynamics of the nebular flow (see section \ref{sec:CharacteristicsOfTurbulentMagneticField}).
We simply assume that $\tau_{\rm diss}$ is a constant in the proper frame as well as $\tau_{\rm conv}$.
Note that $\tau_{\rm diss}$ heats up the plasma and can change the synchrotron cooling rate (see section \ref{sec:RadiativeProperties}).

\subsection{Useful measures of system properties}\label{sec:UsefulMeasuresOfSystemProperties}

We name the first three terms in the right-hand side of equation (\ref{eq:DifferentialEquationForVelocity}) adiabatic, cooling, and conversion terms, respectively.
The last term in the right-hand side of equation (\ref{eq:DifferentialEquationForVelocity}) is zero for $\hat{\Gamma} = 4/3$ and then we do not use dissipation term in the present paper.
For the later convenience, we introduce the ratios of the cooling to adiabatic terms $\xi_{\rm cool}$, and of the conversion to adiabatic terms $\xi_{\rm conv}$.
For example, one of the characteristic ratio with respect to the cooling term is written as
\begin{eqnarray}\label{eq:CoolingRatio}
	\xi_{\rm cool}
	& = &
	\frac{\Lambda_{\rm rad} r}{2 u c} \frac{\hat{\Gamma} - 1}{\hat{\Gamma} p + (2/9)\delta b^2},
\end{eqnarray}
and the cooling term controls the nebular dynamics when $\xi_{\rm cool} > 1$.

We introduce the advection time-scale
\begin{eqnarray}\label{eq:AdvectionTime}
	t_{\rm adv}
	& = &
	\int^{r_{\rm PWN}}_{\rm r_{\rm TS}} \frac{d r}{\beta(r) c},
\end{eqnarray}
where $r_{\rm PWN}$ is the radius of a PWN.
This time-scale is used as a guide of the age of the system in this steady state model.

We also introduce the local synchrotron luminosity per unit length,
\begin{eqnarray}\label{eq:LocalSynLuminosity}
	l_{r,{\rm syn}}(r)
	& \equiv &
	4 \pi r^2 \gamma n P_{\rm syn}
	=
	\kappa_{\rm w} \dot{N}_{\rm GJ} \frac{P_{\rm syn}}{\beta c}.
\end{eqnarray}
From equation (\ref{eq:TotalEnergyConservation}), the synchrotron luminosity radiated away from a PWN of $r_{\rm TS} \le r \le r_{\rm PWN}$ is
\begin{eqnarray}\label{eq:SynLuminosity}
	L_{\rm syn}
	=
	\int^{r_{\rm PWN}}_{r_{\rm TS}} - l_{r,{\rm syn}}(r) d r
	=
	L(r_{\rm TS}) - L(r_{\rm PWN}).
\end{eqnarray}
We define the total synchrotron efficiency as
\begin{eqnarray}\label{eq:SynEfficiency}
	\eta_{\rm syn}
	& \equiv &
	\frac{L_{\rm syn}}{L_{\rm spin}}.
\end{eqnarray}
The total synchrotron efficiency of the Crab Nebula is about 30 \% \citep[][]{Hester08}, although equation (\ref{eq:SynCoolingRate}) is too simple to constrain the model parameters (see section \ref{sec:BroadbandRadiation}).

\section{Application to the Crab Nebula}\label{sec:Application}

Equations (\ref{eq:InternalEnergyConservation}) -- (\ref{eq:TurbulentMagneticEnergyConservation}) and (\ref{eq:DifferentialEquationForVelocity}) are solved by the forth-order Runge-Kutta method.
We consider a PWN of $r_{\rm TS} \le r \le r_{\rm PWN}$.
As discussed in section \ref{sec:InnerBoundaryOfNebulaFlow}, the boundary conditions at $r_{\rm TS}$ are completely characterized by the two parameters, $\sigma_{\rm w}$ and $\kappa_{\rm w}$, for a given $L_{\rm spin}$ (equation (\ref{eq:WindProperties})).
Considering the Crab Nebula, we fix $L_{\rm spin} = 4.6 \times 10^{38}~{\rm erg~s^{-1}}$ and $r_{\rm PWN} = 2.0$ pc throughout this paper.
On the present study, we have two parameters ($\tau_{\rm conv},\tau_{\rm diss}$) and three boundary conditions ($\sigma_{\rm w},\kappa_{\rm w},r_{\rm TS}$).

We focus on the dynamics of the PWN by setting the velocity at the outer boundary as the observed value $\beta(r_{\rm PWN}) c \sim v_{\rm PWN} = 1500~{\rm km~s^{-1}}$ \citep[e.g.,][]{Hester08}.
In section \ref{sec:FlowDynamics}, we will see that $\sigma_{\rm w}$ and $\tau_{\rm conv}$ mainly control the dynamics and we look for the values of $\tau_{\rm conv}$ reproducing the observed $v_{\rm PWN}$ for a given $\sigma_{\rm w}$.
In section \ref{sec:RadiativeProperties}, we will see that $\kappa_{\rm w}$ and $\tau_{\rm diss}$ slightly change the velocity profile through the cooling effect.
In other words, these two parameters change $l_{r,{\rm syn}}$ and $\eta_{\rm syn}$ (see also the discussion in section \ref{sec:BroadbandRadiation}).
We also study dependence of the nebular flow profiles on $r_{\rm TS}$ in section \ref{sec:rTSDependence} as an interesting topic.

We take $\kappa_{\rm w} = 10^4$ and $r_{\rm TS} = 0.1$ pc as fiducial values from past studies \citep[e.g.,][]{Kennel&Coroniti84a, Kennel&Coroniti84b}.
We vary $\tau_{\rm diss}$ from 10 years (fast dissipation) to $\infty$ (no dissipation).
For the wind magnetization parameter $\sigma_{\rm w}$, we study three representative cases of $\sigma_{\rm w} = 0.1, 10, 10^3$ and also $\sigma_{\rm w} = 0.003$ which corresponds to KC model.
Note that the cases of $\sigma_{\rm w} = 1$ do not show a special behavior and we do not study the cases of $\sigma_{\rm w} \ge 10^4$ for which the strong shock jump condition is not available (see equation (\ref{eq:app:StrongShockCondition})).
The values of $\tau_{\rm conv}$ are derived for each combination of $(\sigma_{\rm w}, \kappa_{\rm w}, r_{\rm TS}, \tau_{\rm diss})$ in order to reproduce $\beta(r_{\rm PWN}) c = 1500~{\rm km~s^{-1}}$.

\subsection{Flow dynamics}\label{sec:FlowDynamics}

%
\begin{figure}
 	\includegraphics[width=\columnwidth]{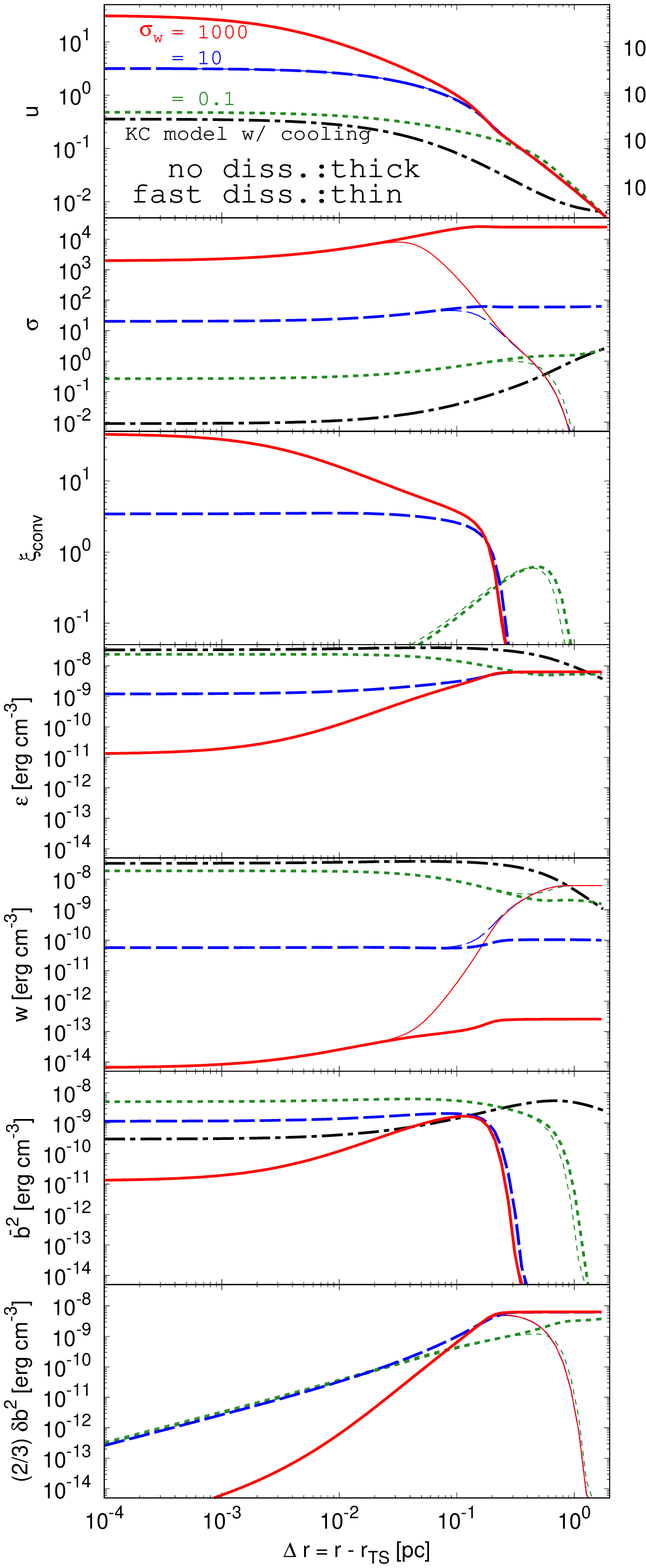}
\caption{
	The nebular flow profiles which satisfy $v_{\rm PWN} = 1500~{\rm km~s^{-1}}$ are plotted.
	The horizontal axis is the distance from the termination shock $\Delta r = r - r_{\rm TS}$.
	The adopted parameters are summarized in Table \ref{tbl:Parameters}.
	$\tau_{\rm diss}$ is 10 years for the thin and $\infty$ for the thick lines and $\sigma_{\rm w}$ is $10^3$ (red solid lines), 10 (blue dashed lines), and 0.1 (green dotted lines).
	For reference, KC model with the cooling effect are plotted in the black dot-dashed line.
}
\label{fig:Dynamics}
\end{figure}
%

%
\begin{table}
	\centering
	\caption{
	Summary of the parameters and the boundary values for Figs. \ref{fig:Dynamics} -- \ref{fig:ConversionModels}.
    }
\label{tbl:Parameters}
\begin{tabular}{cccc|ccc}
	\multicolumn{4}{c}{Adopted} &
	\multicolumn{3}{c}{Derived} \\
	\hline
	$\sigma_{\rm w}$      &
	$\kappa_{\rm w}$      &
	$r_{\rm TS}$[pc]      &
	$\tau_{\rm diss}$[yr] &
	$\tau_{\rm conv}$[yr] &
	$t_{\rm adv}$[kyr]    &
	$\eta_{\rm syn}$      \\
\hline
	\multicolumn{7}{c}{Fig. \ref{fig:Dynamics}} \\
\hline
	$10^3$ & $10^4$ & 0.1   & 10       & 0.228  & 0.434 & 0.0260      \\
	$10$   &        &       & 10       & 0.308  & 0.433 & 0.0259      \\
	$0.1$  &        &       & 10       & 6.51   & 0.435 & 0.0370      \\
	$10^3$ &        &       & $\infty$ & 0.223  & 0.435 & $< 10^{-3}$ \\
	$10$   &        &       & $\infty$ & 0.301  & 0.434 & $< 10^{-3}$ \\
	$0.1$  &        &       & $\infty$ & 7.56   & 0.404 & 0.167       \\

\hline
	\multicolumn{7}{c}{Fig. \ref{fig:Synchrotron}} \\
\hline
	$10^3$ & $10^4$ & 0.1   & 100      & 0.264  & 0.397 & 0.184       \\
	$10$   & $10^4$ &       &          & 0.357  & 0.397 & 0.184       \\
	$0.1$  & $10^4$ &       &          & 7.72   & 0.407 & 0.185       \\
	$10^3$ & $10^2$ &       &          & 0.544  & 0.209 & 0.704       \\
	$10$   & $10^2$ &       &          & 0.752  & 0.209 & 0.705       \\
	$0.1$  & $10^2$ &       &          & 14.0   & 0.230 & 0.775       \\

\hline
	\multicolumn{7}{c}{Fig. \ref{fig:rTS}} \\
\hline
	$10^3$ & $10^4$ & 0.1   & 300      & 0.264  & 0.381 & 0.186       \\
	$10$   &        & 0.1   &          & 0.358  & 0.380 & 0.186       \\
	$0.1$  &        & 0.1   &          & 8.29   & 0.379 & 0.241       \\
	$10^3$ &        & $10^{-3}$ &      & 0.441  & 0.380 & 0.185       \\
	$10$   &        & $10^{-3}$ &      & 0.643  & 0.380 & 0.185       \\
	$0.1$  &        & $10^{-3}$ &      & 14.3   & 0.409 & 0.376       \\
\hline
	\multicolumn{7}{c}{Case A of Fig. \ref{fig:ConversionModels}} \\
\hline
	$10^3$ & $10^4$ & 0.1   & $10^3$   & 0.238  & 0.411 & 0.0701      \\
	$10$   &        &       &          & 0.322  & 0.411 & 0.0755      \\
	$0.1$  &        &       &          & 7.94   & 0.389 & 0.208       \\
\hline
	\multicolumn{7}{c}{Case B of Fig. \ref{fig:ConversionModels}} \\
\hline
	$10^3$ & $10^4$ & 0.1   & $10^3$   & 0.0751 & 0.412 & 0.0707      \\
	$10$   &        &       &          & 0.200  & 0.411 & 0.0755      \\
	$0.1$  &        &       &          & 7.84   & 0.389 & 0.208       \\
\hline
	\multicolumn{7}{c}{KC model} \\
\hline
	$0.003$ & $10^4$ & 0.1 & $\infty$ & $\infty$ & 0.637 & 0.359 \\
\end{tabular}
\end{table}

Fig. \ref{fig:Dynamics} compares the flow dynamics for different dissipation time-scales $\tau_{\rm diss}$.
All three magnetization cases of $\sigma_{\rm w} = 0.1$ (red solid), $10$ (blue dashed), $10^3$ (green dotted) are over-plotted.
The thin and thick lines correspond to the fast ($\tau_{\rm diss} = $10 yr) and no ($\tau_{\rm diss} = \infty$) dissipation models, respectively.
$r_{\rm TS} = 0.1$ pc, and $\kappa_{\rm w} = 10^4$ are common in all these plots.
For reference, we plot the profiles of KC model including the synchrotron cooling effect ($\sigma_{\rm w} = 0.003$, $\kappa_{\rm w} = 10^4$, $r_{\rm TS} = 0.1$ pc and $\tau_{\rm diss} = \tau_{\rm conf} = \infty$).
Note that the horizontal axes are not the distance from the pulsar $r$ but from the termination shock $\Delta r \equiv r - r_{\rm TS}$.
The used parameters are summarized in Table \ref{tbl:Parameters}.

The thin and thick lines are overlapped for the radial velocity $u(r)$ (top panel), the total enthalpy density $\epsilon(r)$ (forth panel) and the toroidal magnetic field $\bar{b}^2(r)$ (sixth panel) profiles of the nebular flow.
These lines demonstrate that $\tau_{\rm diss}$ is not important for the flow velocity profile.
The derived conversion time-scales $\tau_{\rm conv}$ are also almost independent from $\tau_{\rm diss}$ and we conclude that the flow dynamics is controlled mainly by $\tau_{\rm conv}$ and not by $\tau_{\rm diss}$ for a given $\sigma_{\rm w}$.

The difference between the thin and thick lines are apparent for the profiles of the magnetization $\sigma(r)$ (second panel), the enthalpy density $w$ (fifth panel) and the turbulent magnetic field $\delta b^2$ (bottom panel).
These differences caused by $\tau_{\rm diss}$ affect the radiative properties which will be discussed in the next section \ref{sec:RadiativeProperties}.
Note that we do not need to reduce $\sigma(r)$ in order to decelerate the flow (the thick lines in the second panel of Fig. \ref{fig:Dynamics}).

For the cases of a high-$\sigma_{\rm w}$ wind ($\sigma_{\rm w} = 10$ and $10^3$), the velocity is still relativistic behind the termination shock $u(r_{\rm TS}) \approx \sqrt{\sigma_{\rm w}} > 1$ (equation (\ref{eq:app:HighSigmaBoundaryVelocity})) and, as is evident from the third panel of Fig. \ref{fig:Dynamics}, the conversion term $\xi_{\rm conv}$ plays a role for gradual deceleration to a non-relativistic velocity by the flow getting to $\Delta r \sim 0.2$ pc.
After conversion of the toroidal to turbulent magnetic field, the flow behaves as the hydrodynamic post-shock flow of $u \propto r^{-2}$ because the turbulent magnetic field behaves as the relativistic gas (see section \ref{sec:CharacteristicsOfTurbulentMagneticField}).
We also studied a low-$\sigma_{\rm w}$ wind of $\sigma_{\rm w} = 0.1$, which is still a much larger magnetization than KC model of $\sigma_{\rm w} = 0.003$.
For the case of a low-$\sigma_{\rm w}$ wind, we require a finite $\xi_{\rm conv}$ but $\lesssim 1$ to decelerate the flow to $1500~{\rm km~s^{-1}}$.

For another measure of the velocity profiles, we tabulated the advection time-scale $t_{\rm adv}$ in Table \ref{tbl:Parameters}.
$t_{\rm adv}$ is almost independent from $\sigma_{\rm w}$, i.e., the velocity profiles at $\Delta r \lesssim 0.1$ pc does not contribute to $t_{\rm adv}$.
Although $t_{\rm adv}$ is a bit smaller than the age of the Crab Nebula ($t_{\rm age} \sim$ kyr) for all the cases including KC model, three-dimensional turbulent flow structures beyond $\Delta r \gtrsim 1$ pc would resolve this discrepancy in practice \citep[e.g.,][]{Porth+14b}.

The forth panel of Fig. \ref{fig:Dynamics} shows the profiles of the total enthalpy density $\epsilon(r)$.
The synchrotron cooling hardly changes $\epsilon(r)$ and the total energy flux $L(r)$ is almost conserved because $\eta_{\rm syn} \ll 1$.
From equation (\ref{eq:TotalEnergyFlux}), the Lorentz contraction and the Doppler effect lower $\epsilon(r)$ at the region of $u(r) \gg 1$ for the high-$\sigma_{\rm w}$ cases.
$\epsilon(r_{\rm PWN})$ is the same for all the models including KC model because $v_{\rm PWN} = 1500~{\rm km~s^{-1}}$ is common for all of them.
The total pressure at $r_{\rm PWN}$ will also be almost the same for all the models.
The pressure balance condition between the nebular flow and the outer supernova ejecta is not affected by $\tau_{\rm diss}$ and would be satisfied for the high-$\sigma_{\rm w}$ cases.

For the high-$\sigma_{\rm w}$ cases, the enthalpy density $w$ slightly increases with $r$ even without magnetic dissipation (the thick red solid and thick blue dashed lines in the fifth panel of Fig. \ref{fig:Dynamics}).
Because the post-shock flow is hot, equation (\ref{eq:InternalEnergyConservation}) without the right-hand side terms gives $w \sim 4 p \propto n^{4/3} \propto (u r^2)^{-4/3}$.
$w$ increases with $r$ when $u$ decreases faster than $\propto r^{-2}$.
The flow is heavily decelerated when $\xi_{\rm conv} > 1$, and then the flow is adiabatically heated up rather than cooled by the synchrotron radiation.
Plasma heating by magnetic dissipation is much more effective for the high-$\sigma_{\rm w}$ cases as seen in the differences between the thin and thick lines.
Even for the low-$\sigma_{\rm w}$ case, $w$ slightly increases by magnetic dissipation at $\Delta r \gtrsim 0.3$ pc.

The sixth and bottom panels of Fig. \ref{fig:Dynamics} shows the profiles of $\bar{b}^2(r)$ and $\delta b^2(r)$, respectively.
When the right-hand side of equation (\ref{eq:ToroidalMagneticEnergyConservation}) is neglected, the profiles of $\bar{b}^2$ is described by the conservation of the magnetic flux $\bar{b} \propto (u r)^{-1}$.
For the high-$\sigma_{\rm w}$ cases, because of the magnetic flux conservation, $\bar{b}^2$ increases with $r$ at $\Delta r < 0.1$ pc.
Beyond $\Delta r > 0.1$ pc, $\bar{b}^2$ decreases rapidly with $r$ by conversion of $\bar{b}^2$ to $\delta b^2$.

$\delta b^2$ is initially zero and increases with $r$ by conversion from $\bar{b}^2$.
The profiles of $\delta b^2$ is the same as the relativistically hot plasma $\delta b^2 \propto (u r^2)^{-4/3}$ when we neglect the right-hand side of equation (\ref{eq:TurbulentMagneticEnergyConservation}).
We see such a behavior from the thick red solid and thick blue dashed lines in the bottom panel of Fig. \ref{fig:Dynamics} at $\Delta r > 0.2$ pc, i.e., $\delta b^2 =$ const. for $u \propto r^{-2}$.

\subsection{Implications for radiation}\label{sec:RadiativeProperties}

%
\begin{figure}
 	\includegraphics[width=\columnwidth]{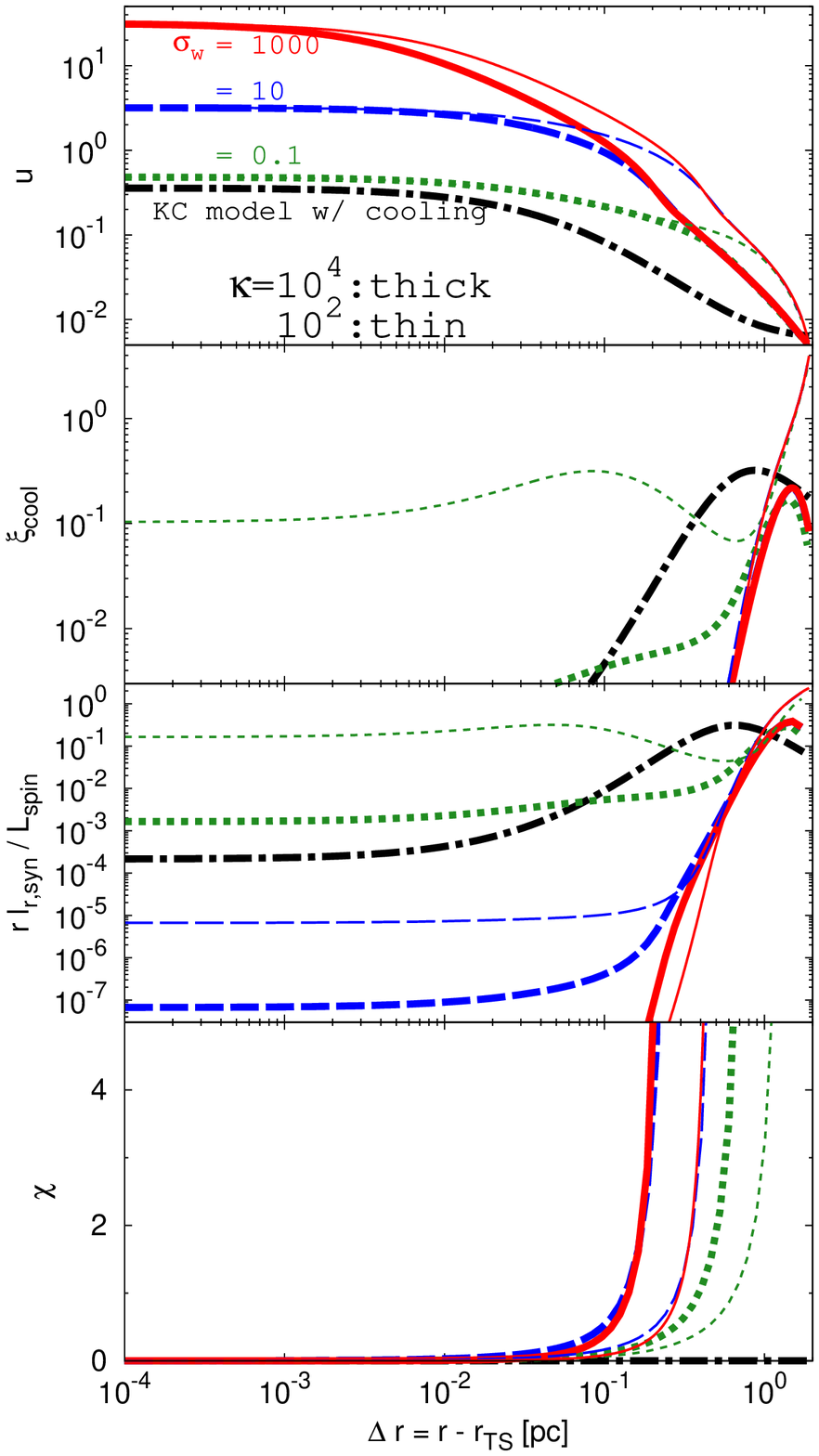}
\caption{
	The resultant flow profiles as a function of $\Delta r$.
	Adopted parameters are summarized in Table \ref{tbl:Parameters} and combinations of ($\sigma_{\rm w}, \kappa_{\rm w}$) are different for each line, where the thin and thick lines are $\kappa_{\rm w} = 10^2$ and $10^4$, respectively.
}
\label{fig:Synchrotron}
\end{figure}

Although our model of the synchrotron cooling is too simple to account for the observed spectral features (see section \ref{sec:BroadbandRadiation} for details), it is still enough to understand effects of the synchrotron cooling on the flow dynamics.
In Fig. \ref{fig:Synchrotron}, we set $\tau_{\rm diss} =$ 100 yr and the pair multiplicity $\kappa_{\rm w}$ is different between the thin ($\kappa_{\rm w} = 10^2$) and thick ($\kappa_{\rm w} = 10^4$) lines.
The used parameters are summarized in Table \ref{tbl:Parameters}.
Note that $\kappa_{\rm w} = 10^2$ is just demonstrations and is not realistic cases considering the Crab Nebula \citep[e.g.,][]{Tanaka&Asano17}.

It is evident from Table \ref{tbl:Parameters} that the synchrotron cooling is more efficient for the cases of $\kappa_{\rm w} = 10^2$ than those of $\kappa_{\rm w} = 10^4$, i.e., $\eta_{\rm syn}$ is larger for a smaller $\kappa_{\rm w}$.
Although the number of the particles emitting the synchrotron radiation is less for smaller $\kappa_{\rm w}$, the temperature is larger for smaller $\kappa_{\rm w}$ as $\gamma_{\rm th} \propto \kappa^{-1}_{\rm w}$ at immediate downstream of the termination shock.
It is clear that the synchrotron cooling is less important for larger $\kappa_{\rm w}$.

Dependence of $\eta_{\rm syn}$ on $\tau_{\rm diss}$ is a bit more complicated than that on $\kappa_{\rm w}$.
For the high-$\sigma_{\rm w}$ cases, the moderate dissipation time-scale of a few hundreds years gives a maximum value of $\eta_{\rm syn}$.
Setting $\kappa_{\rm w} = 10^4$, the synchrotron efficiencies $\eta_{\rm syn}$ are larger for $\tau_{\rm diss} = 100$ yr (Fig. \ref{fig:Synchrotron}) than those for $\tau_{\rm diss} = 10$ yr and $\infty$ (Fig. \ref{fig:Dynamics}), e.g., we have $(\eta_{\rm syn}, \tau_{\rm diss}) \sim (0.026, 10~{\rm yr})$, $(0.18, 100~{\rm yr})$ and $(< 10^{-3}, \infty)$ for the high-$\sigma_{\rm w}$ cases.
For no magnetic dissipation $\tau_{\rm diss} = \infty$, plasma has no enough thermal energy $\gamma_{\rm th}$ to radiate.
On the other hand, for fast dissipation $\tau_{\rm diss} = 10$ yr, the magnetic energy is quickly converted to the plasma thermal energy but the synchrotron radiation is inefficient because of the small magnetic energy.
Taking the results with different values of $\tau_{\rm diss}$ (Figs. \ref{fig:rTS} and \ref{fig:ConversionModels}) in advance, $\eta_{\rm syn}$ would be maximized at $\tau_{\rm diss} \sim t_{\rm adv}$ a few hundreds years for the high-$\sigma_{\rm w}$ cases.

The velocity profiles (top panel) slightly change for the strong cooling regime ($\kappa_{\rm w} = 10^2$ and $\tau_{\rm diss} = 100$ yr).
$t_{\rm adv}$ for the cases of $\kappa_{\rm w} = 10^2$ is about a factor of two smaller than those of $\kappa_{\rm w} = 10^4$.
For $\kappa_{\rm w} = 10^2$, the second panel of Fig. \ref{fig:Synchrotron} shows that $\xi_{\rm cool}$ plays a role beyond $\sim 1$ pc and thus the flow velocity drops more rapidly than $u \propto r^{-2}$.
The synchrotron cooling hardly affects the flow dynamics ($\xi_{\rm cool} < 1$) everywhere behind the shock for $\kappa_{\rm w} \ge 10^4$.

The third panel of Fig. \ref{fig:Synchrotron} shows the profiles of the normalized local synchrotron luminosity $r l_{r,{\rm syn}} / L_{\rm spin}$.
The local synchrotron luminosity $r l_{r,{\rm syn}}$ has a peak at the radius satisfying $\sigma(r) \sim 1$, because $P_{\rm syn}$ is the product of the particle and magnetic energy densities (equation (\ref{eq:SynCoolingRate})).
$\tau_{\rm diss}$ of a few hundreds years corresponds to the cases that $\sigma \sim 1$ at around $r_{\rm PWN}$ (see the third row of Fig. \ref{fig:Synchrotron}) and thus the synchrotron efficiency $\eta_{\rm syn}$ is high in those cases.
As is found from the thin lines in Fig. \ref{fig:Dynamics} ($\tau_{\rm diss} = 10$ yr), $\sigma(r)$ becomes order unity and $r l_{r,{\rm syn}}$ has peak at the inner nebula $\Delta r \sim 0.5$ pc, at which KC model also shows the peak $r l_{r,{\rm syn}}$.

The local synchrotron luminosity $l_{r,{\rm syn}}$ is larger for a smaller $\kappa_{\rm w}$.
For the high-$\sigma_{\rm w}$ cases, $l_{r,{\rm syn}}$ at close to $r_{\rm TS}$ is fairly small and rapidly increases with $r$ at $\Delta r \gtrsim 0.1$ pc due to the plasma heating by magnetic dissipation and to flow deceleration (see equation (\ref{eq:LocalSynLuminosity}).
The behaviors at the immediate downstream of the termination shock are consistent with $l_{r,{\rm syn}} \propto \kappa_{\rm w}^{-1} \sigma^{-4}_{\rm w}$ obtained from equations (\ref{eq:app:SigmaWind}) -- (\ref{eq:app:HighSigmaBoundaryPressure}), where $\kappa_{\rm w} \gamma_{\rm th} \sim p_{\rm TS} / m_{\rm e} c^2 \propto \sigma^{-3/2}_{\rm w}$ and $\bar{b}^2 \sim b^2_{\rm TS} \propto \sigma^{-1}_{\rm w}$ for $1 \ll \sigma_{\rm w} \ll 10^4$.

The bottom panel of Fig. \ref{fig:Synchrotron} shows the profiles of randomness of the magnetic field $\chi$.
The magnetic field should be tangled in order to decelerate the flow and $\chi \gg 3$ at the brightest region of the nebula $\Delta r > 1$ pc.
Although the synchrotron emission from the Crab Nebula is highly polarized ($\sim 20 - 30 \%$) \citep[e.g.,][]{Bietenholz&Kronberg91, Aumont+10, Moran+13, Chauvin+17}, we do not observe polarized synchrotron emission in the present formulation (see discussion in section \ref{sec:Comparison}).

\subsection{Termination shock radius}\label{sec:rTSDependence}

%
\begin{figure}
 	\includegraphics[width=\columnwidth]{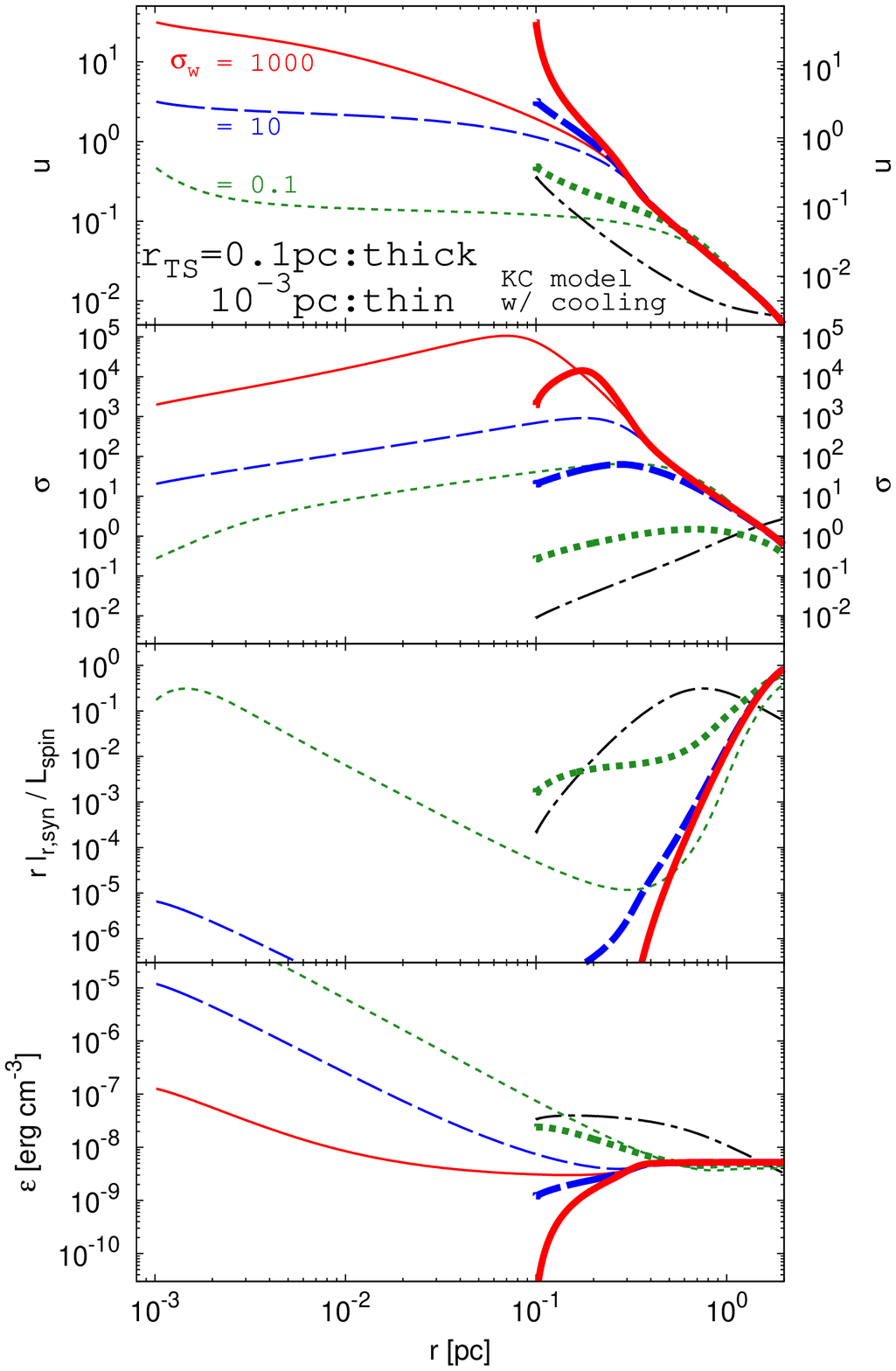}
\caption{
	The resultant flow profiles as a function of $r$ ($\ne \Delta r$).
	Adopted parameters are summarized in Table \ref{tbl:Parameters} and combinations of ($\sigma_{\rm w}, r_{\rm TS}$) are different for each line, where the thin and thick lines are $r_{\rm TS} = 10^{-3}$ pc and $0.1$ pc, respectively.
}
\label{fig:rTS}
\end{figure}
%

Here, we study the cases that the termination shock radius is much smaller than the customarily used value of $r_{\rm TS} = 0.1$ pc.
$r_{\rm TS} = 0.1$ pc is inferred from the optical `wisps' \citep[][]{Scargle69} and also the X-ray `inner ring' \citep[][]{Weisskopf+00}.
However, the origin of these inner structures is theoretically unclear \citep[c.f.,][]{Hoshino+92} and synthetic synchrotron maps based on the multi-dimensional ideal MHD simulations do not fully reproduce these features \citep[see section 5 of][]{Porth+14a}.
An independent argument for $r_{\rm TS} \sim 0.1$ pc has done by \cite{Rees&Gunn74} and also \cite{Kennel&Coroniti84a}.
They had discussions on the dynamics of the Crab Nebula (expansion velocity and the pressure balance with supernova ejecta) assuming ideal MHD so that the result would be different with the use of our formulation (see also section \ref{sec:TerminationShock}).

Fig. \ref{fig:rTS} shows the flow profiles as a function of the distance from the pulsar $r$ and not from the termination shock $\Delta r$ (Figs. \ref{fig:Dynamics} and \ref{fig:Synchrotron}).
The termination shock radius $r_{\rm TS}$ is different between the thin ($10^{-3}$ pc) and thick ($0.1$ pc) lines.
$\kappa_{\rm w} = 10^4$ and $\tau_{\rm diss} = 300$ yr are common for all the lines.
The other parameters are summarized in Table \ref{tbl:Parameters}.

In the top panel of Fig. \ref{fig:rTS}, we find that both the thin and thick lines can satisfy $v_{\rm PWN} = 1500~{\rm km~s^{-1}}$ again.
The values of $\tau_{\rm conv}$ are only about a factor of two larger for the cases of $r_{\rm TS} = 10^{-3}$ pc than those of $r_{\rm TS} = 0.1$ pc.
If we rewrite the four velocity $u$ as a function of $\Delta r$, we find that the thin and thick lines are almost overlapped because the characteristic length-scale of the flow dynamics is only $u(r_{\rm TS}) c \tau_{\rm conv}$.
Note that both $u(r_{\rm TS})$ and $\sigma(r_{\rm TS})$ are independent from $r_{\rm TS}$ (equations (\ref{eq:app:RHConditionVelocity}) and (\ref{eq:app:RHConditionSigma})).

As we have already studied in section \ref{sec:RadiativeProperties}, a peak of the local synchrotron luminosity (third panel) is around the radius satisfying $\sigma(r) \sim 1$ (second panel).
The characteristic length-scale of the radiation is $u(r_{\rm TS}) c \tau_{\rm diss}$ and is independent from $r_{\rm TS}$ again.
Especially for the high-$\sigma_{\rm w}$ cases, the immediate downstream of the termination shock is underluminous and both $t_{\rm adv}$ and $\eta_{\rm syn}$ are similar for different $r_{\rm TS}$.
If we observe the surface brightness profiles of the third panel of Fig. \ref{fig:rTS}, we would not distinguish the thin and thick lines for the high-$\sigma_{\rm w}$ cases.
The local synchrotron luminosity at $r_{\rm TS}$ is larger for smaller $r_{\rm TS}$ because the energy densities of both the plasma and the magnetic field decrease with radius (the bottom panel of Fig. \ref{fig:rTS}), i.e., $r_{\rm TS}$ should not be arbitrarily small in order not to be larger $\eta_{\rm syn}$ than the observations.
For example, $r_{\rm TS} \gtrsim 10^{-3}$ pc for $\sigma_{\rm w} = 10^{-1}$.

Bottom panel of Fig. \ref{fig:rTS} is the radial distributions of the total enthalpy density $\epsilon(r)$.
The total enthalpy density at $r_{\rm TS}$ strongly depend on $r_{\rm TS}$ as $\epsilon(r_{\rm TS}) \sim p_{\rm TS} (1 + \sigma_{\rm TS}) \propto (r^2_{\rm TS} \sigma_{\rm w})^{-1}$ from equations (\ref{eq:app:SigmaWind}) -- (\ref{eq:app:HighSigmaBoundaryPressure}).
On the other hand, the total enthalpy density at $r_{\rm PWN}$ are almost the same because of the inefficient radiative cooling $L(r_{\rm PWN}) \sim L_{\rm spin}$ and the same $u(r_{\rm PWN})$ as already described in section \ref{sec:FlowDynamics}.
In other words, the pressure balance condition at the contact discontinuity of the PWN and the supernova ejecta is also independent of $r_{\rm TS}$.

\section{Discussion}\label{sec:Discussion}

%
\begin{figure}
 	\includegraphics[width=\columnwidth]{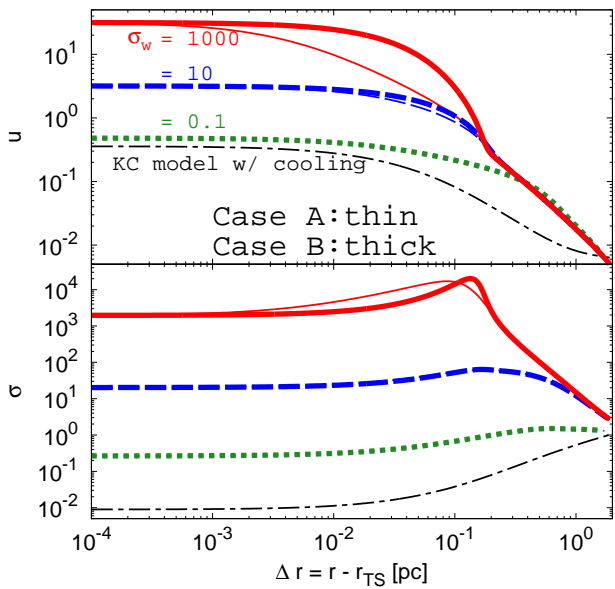}
\caption{
	The nebular flow profiles for cases A (thin) and B (thick), where the conversion time-scale is constant in the proper frame for case A while in the observer frame for case B.
	The adopted parameters are summarized in Table \ref{tbl:Parameters}.
}
\label{fig:ConversionModels}
\end{figure}
%

\subsection{Comparison with other studies}\label{sec:Comparison}

Our results demonstrate that the disappearance of the toroidal magnetic field is essential for the flow deceleration.
Conversion of the toroidal to the turbulent magnetic field is our approach, while \citet{Zrake&Arons17} considered dissipation of the toroidal magnetic field directly into the heat of plasma following \citet{Lyubarsky&Kirk01} and \citet{Drenkhahn02}.
In order to compare the flow dynamics with the previous studies, their dissipation time-scale is corresponding to our conversion time-scale $\tau_{\rm conv}$.
Only in this section \ref{sec:Comparison}, we use the term `disappearance time-scale' to express for the both time-scales which are almost the same in the dynamical point of view.
Our dissipation-time scale $\tau_{\rm diss}$ will be discussed in the next section \ref{sec:BroadbandRadiation}.

Although we adopted the disappearance time-scale is a constant in the proper frame, the other studies took different ways.
For example, \citet{Lyubarsky&Kirk01} and \citet{Drenkhahn02} considered the annihilation of the anti-parallel toroidal magnetic field in terms of the flow acceleration \citep[c.f.,][]{Coroniti90} and the time-scale is set to constant in the observer (pulsar) frame.
In Fig. \ref{fig:ConversionModels}, we compare the flow profiles for the cases that the disappearance time-scale is a constant in the proper (case A) and the observer frame (case B).
The parameters are tabulated in Table \ref{tbl:Parameters}.
Case A corresponds to our study, while we rewrite $\tau_{\rm conv} \rightarrow \gamma \tau_{\rm conv}$ for case B.
They show qualitatively the same behaviors and differences between cases A and B are obvious only for the cases of $\sigma_{\rm w} = 10^3$ because $u(r) \gg 1$ behind the termination shock.

Another choice of the disappearance time-scale is adopted by \citet{Zrake&Arons17}.
They considered the turbulent relaxation model, in which the disappearance time-scale changes with the density of the plasma and the magnetic field strength, i.e., the time-scale is not constant with $r$.
However, the disappearance time-scale is almost constant in the proper frame at close to $r_{\rm TS}$ and their value of $\sim 0.3$ yr is consistent with $\tau_{\rm conv}$ of our study for the high-$\sigma_{\rm w}$ cases \citep[see the bottom panel in Fig. 4 of][]{Zrake&Arons17}.
Note that although the disappearance time-scale increases with $r$ for the turbulent relaxation model, the time-scale at close to $r_{\rm TS}$ is important for the flow deceleration because we have $\xi_{\rm conv} > 1$ only at close to $r_{\rm TS}$ (see Fig. \ref{fig:Dynamics}).

\subsection{Broadband radiation properties}\label{sec:BroadbandRadiation}

Although we included the cooling by the synchrotron radiation in equations (\ref{eq:TotalEnergyConservation}) and (\ref{eq:InternalEnergyConservation}), we do not take account for the broadband spectral properties of the Crab Nebula in this paper.
It is important to develop the model accounting for the broadband spectrum based on the present model of the nebular dynamics.
Although the dynamics of the Crab Nebula is explained for $\sigma(r) \gg 1$ (e.g., no dissipation cases in Fig. \ref{fig:Dynamics}), the past one-zone spectral studies of the Crab Nebula showed that spatially averaged value of $\sigma(r)$ should be much smaller than unity in order to reproduce the synchrotron to inverse Compton flux ratio \citep[e.g.,][]{Tanaka&Takahara10}.
A finite value of $\tau_{\rm diss}$ is required in view of the nebula spectrum.

However, the treatment of the dissipation time-scale in the broadband spectral model is not trivial because the plasma heating and the particle acceleration mechanisms by magnetic dissipation is not well understood.
Almost all the spectral studies of the Crab Nebula considered that the particles are accelerated at the termination shock and are just cooled radiatively and also adiabatically along with the flow \citep[c.f.,][]{Kennel&Coroniti84b, Porth+14a}.
On the other hand, especially for the high-$\sigma_{\rm w}$ cases of the present study, the particles should be re-accelerated inside the PWN in order to reproduce the observed $\eta_{\rm syn}$ and then the particle acceleration at the termination shock may not be required \citep[c.f.,][]{Begelman&Kirk90, Sironi&Spitkovsky09}.
The stochastic acceleration model by \citet{Tanaka&Asano17} would be one possible treatment of the re-acceleration, while the shock acceleration is also required in their model.
We have to develop the broadband spectral model including magnetic dissipation.

\subsection{Spectrum of Magnetic Turbulence}\label{sec:SpectrumOfTurbulence}

For simplicity, we do not take account of the energy spectrum of the magnetic turbulence in this study.
For example, in terms of the (non-relativistic) solar wind plasma, \citet{Zhou&Matthaeus90a,Zhou&Matthaeus90b,Zhou&Matthaeus90c} studied the transport equation of (locally isotropic) MHD turbulence of spectral energy densities $E_k$, which would be formally written as
\begin{eqnarray}
	\partial_t E_k + \mathcal{L}_{\bm x} E_k + T_k
	=
    C_k - D_k
	.\label{eq:ZM90}
\end{eqnarray}
Equation (\ref{eq:TurbulentMagneticEnergyConservation}) would be closely related with equation (\ref{eq:ZM90}) integrated over wave number $k$, where $\int E_k dk \propto \delta b^2$.
$\mathcal{L}_{\bm x}$ is a linear spatial transport operator acting on $E_k$ (the left-hand side of equation (\ref{eq:TurbulentMagneticEnergyConservation})), $T_k$ represents the energy transport in $k$--space (i.e., a cascade process) satisfying $\int T_k dk = 0$, $C_k$ corresponds to the energy conversion from the ordered magnetic field to the large-scale turbulence ($\int C_k dk \propto \bar{b}^2 / \tau_{\rm conv}$) and $D_k$ is energy dissipation of the small-scale turbulence ($\int D_k dk \propto \delta  b^2 / \tau_{\rm diss}$).

Our model of the transport equation of the turbulent magnetic field (equation (\ref{eq:TurbulentMagneticEnergyConservation})) would be obtained from equation (\ref{eq:ZM90}) by integrating over $k$ with steady approximation $\partial_t = 0$.
Direct dissipation of the toroidal magnetic field considered by \citet{Zrake&Arons17} would correspond to the case that all the left-hand side terms in Equation (\ref{eq:ZM90}) are neglected ($C_k = D_k$) as we have already mentioned in section \ref{sec:BasicEquations}.
For a steady and homogeneous ($\partial_t = \mathcal{L}_{\bm x} = 0$) system, small-scale dissipation has to balance large-scale injection $\int C_k dk = \int D_k dk$, i.e., $\delta b^2 / \tau_{\rm diss} = \bar{b}^2 \tau_{\rm conv}$, and the transport term $T_k$ characterizes the spectrum of inertial range, e.g., $E_k \propto k^{-3/5}$ for the Kolmogorov-like turbulence.
However, because we consider the steady but inhomogeneous $\mathcal{L}_{\bm x} \ne 0$ system, we have $\delta b^2 / \tau_{\rm diss} \ne \bar{b}^2 / \tau_{\rm conv}$ in general.
The energy spectrum of the turbulence is not necessarily as simple as the Kolmogrov-like turbulence even at inertial range.

\subsection{Turbulent magnetic field and polarization}\label{sec:TurbulentComponent}

Introduction of the turbulent component of the magnetic field brings another advantage for polarization properties of PWNe.
Observed polarization properties of the Crab Nebula infer a finite magnitude of the turbulent component of the magnetic field ($\chi > 0$) across the nebula \citep[e.g.,][]{Bucciantini+17}.
However the model of \citet{Zrake&Arons17} has no turbulent magnetic field $\delta b^2 = 0$ everywhere in the nebula.
Although their model changes the profiles of the four velocity $u$ and the local synchrotron luminosity $l_{r,{\rm syn}}$ from KC model, the polarization fraction is the same, i.e., almost the synchrotron maximum value of $\sim 70 \%$ corresponding to the synchrotron emission from the completely ordered magnetic field.
Our model can even address the spatial variation of the randomness $\chi$.
Combined with the broadband spectral modeling discussed in section \ref{sec:BroadbandRadiation}, we will be able to study spatially resolved multi-wavelength polarization properties of the Crab Nebula.


However, the phenomenological conversion term should be modified before further studies of radiation properties of PWNe.
As we have already mentioned in section \ref{sec:RadiativeProperties}, in the high-$\sigma_{\rm w}$ cases, the synchrotron cooling is efficient when the significant amount of the magnetic energy is dissipated.
Since only the turbulent component of the magnetic field can dissipate in our model, the ordered magnetic field hardly exists at the synchrotron emitting region and the observed highly polarized emission is not reproduced from the present formulation.
For example, the ordered magnetic field remains not to be fully converted to the turbulent one by rewriting $\bar{b}^2 / \tau_{\rm conv} \rightarrow \bar{b}^2 (\chi_{\rm max} - \chi) / \tau_{\rm conv}$ \citep[c.f.,][]{Drenkhahn02}.
In order to increase the polarization degree, we have only to introduce the maximum value of the randomness $\chi_{\rm max} \sim 3$.
This kind of modification of the phenomenological terms would be an option or we should try to develop a model of a physically motivated conversion term from, for example, the results of numerical simulations \citep{Porth+14a,Porth+14b}.

\subsection{Pressure balance at termination shock}\label{sec:TerminationShock}

Finally, we discuss the interesting results for the termination shock radius.
We studied the case of $r_{\rm TS} = 10^{-3}$ pc and showed that the observed dynamical characteristics are explained even for $r_{\rm TS} \ll 0.1$ pc.
It seems difficult to constrain $r_{\rm TS}$ only from the dynamical argument in the present model.
The customarily used value of the termination shock radius $r_{\rm TS} = 0.1$ pc comes from the observed size of the inner structures in optical \citep{Scargle69} and in X-rays \citep[][]{Weisskopf+00}, and from the hydrodynamical discussion of the expansion velocity by \citet{Rees&Gunn74}.
We have argued that the latter is not always applicable.
\citet{Rees&Gunn74} considered that the pulsar wind dynamical pressure $L_{\rm spin} / (4 \pi r^2_{\rm TS} c)$ equals to the pressure inside the nebula $L_{\rm spin} t_{\rm age} / (4 \pi r^3_{\rm PWN})$ with $r_{\rm PWN} = v_{\rm PWN} t_{\rm age}$ and thus $r_{\rm TS} / r_{\rm PWN} = \sqrt{v_{\rm PWN} / c} \sim 0.1$.
Although the hydrodynamic post-shock flow is almost spatially isobaric, our nebular flow is allowed to have a spatial gradient of the total pressure, i.e., $\epsilon(r_{\rm TS}) \approx \epsilon(r_{\rm PWN})$ is not always satisfied (the forth panel of Fig. \ref{fig:rTS}).
Because the local synchrotron luminosity at $r_{\rm TS}$ becomes brighter for smaller $r_{\rm TS}$ (the third panel of Fig. \ref{fig:rTS}), broadband brightness profile calculated from the broadband spectral model would constrain the value of $r_{\rm TS}$ \citep[c.f.,][]{Ishizaki+17}.


\section{Conclusions}\label{sec:Conclusions}

In this paper, we extend the one-dimensional spherically symmetric steady model by \citet{Kennel&Coroniti84a} by formulating conversion of the toroidal to turbulent magnetic field and dissipation of the turbulent magnetic field in phenomenological ways.
Disordering of pure toroidal magnetic structure is inferred from the three-dimensional relativistic MHD simulation \citep{Porth+14a} and partly takes into account the multi-dimensional effects, for example, the kink instability inside the nebula \citep{Begelman98}.
Randomization of the magnetic topology would be described within the ideal MHD approximation.
Magnetic dissipation by reconnection is a non-ideal MHD effect requiring a finite resistivity.

Conversion of the toroidal to turbulent magnetic field is more important for the flow deceleration than dissipation of the turbulent magnetic field.
This is consistent with the ideal MHD simulation by \citet{Porth+14a}, and numerical magnetic dissipation in their calculation is not essential for the confinement of a PWN by supernova ejecta.
For any values of the wind magnetization $\sigma_{\rm w}$ much larger than unity, we obtain the observed expansion velocity of the Crab Nebula $v_{\rm PWN} = 1500~{\rm km~s^{-1}}$ by adjusting the conversion time-scale $\tau_{\rm conv}$, which is $\sim 0.3$ yr for the high-$\sigma_{\rm w}$ cases.
We do not need to reduce $\sigma_{\rm w}$ at the far upstream wind region or at the termination shock.

The dissipation time-scale $\tau_{\rm diss}$ changes the synchrotron radiation efficiency $\eta_{\rm syn}$, especially for the high-$\sigma_{\rm w}$ cases.
In the present model of the synchrotron cooling effect, $\tau_{\rm diss}$ of a few hundreds years gives a maximum $\eta_{\rm syn}$.
For the conventional value of the pair multiplicity $\kappa_{\rm w} > 10^4$, the synchrotron cooling effect does not contribute to the flow deceleration for any $\tau_{\rm diss}$.

The present model of the PWN dynamics alone does not constrain $(\sigma_{\rm w}, \kappa_{\rm w}, \tau_{\rm conv}, \tau_{\rm diss})$ and even the termination shock radius $r_{\rm TS}$.
We find that $r_{\rm TS} \ll 0.1$ pc reproduces the dynamical features of the Crab Nebula.
It is important to extend the present model by taking into account the broadband emission.
Especially, the broadband surface brightness profile will be an interesting tool in order to distinguish the model parameters \citep[e.g.,][]{Ishizaki+17}.

\section*{Acknowledgments}

S. J. T. would like to thank Y. Ohira, W. Ishizaki, S. Kimura, K. Asano, R. Yamazaki, and K. Ioka for useful discussion.
S. J. T. would also like to thank anonymous referee for his/her very helpful comments.
This work is supported by JSPS Grants-in-Aid for Scientific Research Nos. 17H18270 (ST), 15H05437 (KT), 15H05440 (NT) and also by a JST grant Building of Consortia for the Development of Human Resources in Science and Technology (KT).




\bibliographystyle{mnras}
\bibliography{draft} 




\appendix

\section{The shock jump condition}\label{app:JumpCondition}

We adopt the shock jump condition of the perpendicular relativistic shock from \citet{Kennel&Coroniti84a}, i.e., the upstream is cold plasma with a pure toroidal magnetic field.
The radiative cooling and the turbulent magnetic field are not considered at the termination shock.
$L_{\rm spin},~\sigma_{\rm w}$, and $\kappa_{\rm w}$ are used to determine the upstream property (see section \ref{sec:InnerBoundaryOfNebulaFlow}).
\citet{Kennel&Coroniti84a} wrote the post-shock properties as
\begin{eqnarray}
	u^2_{\rm TS}
	& = &
	\frac{8 \sigma^2_{\rm w} + 10 \sigma_{\rm w} + 1}{16 (\sigma_{\rm w} + 1)} \nonumber \\
    & + & \frac{\sqrt{64 \sigma^2_{\rm w} (\sigma_{\rm w} + 1)^2 + 20 \sigma_{\rm w} (\sigma_{\rm w} + 1) + 1}}{16 (\sigma_{\rm w} + 1)}
	, \label{eq:app:RHConditionVelocity} \\
	\frac{n_{\rm TS}}{n_{\rm w}}
	& = &
	\frac{b_{\rm TS}}{b_{\rm w}}
	=
	\frac{u_{\rm w}}{u_{\rm TS}}
	, \label{eq:app:RHConditionDensity} \\
	p_{\rm TS}
	& = &
	\frac{n_{\rm w} m_{\rm e} c^2 u^2_{\rm w}}{4 u_{\rm TS} \gamma_{\rm TS}}
	\left[
		1 + \sigma_{\rm w} (1 - \beta^{-1}_{\rm TS})
	\right]
	, \label{eq:app:RHConditionPressure} \\
	\sigma_{\rm TS}
	& \equiv &
	\frac{\bar{b}^2_{\rm TS}}{w_{\rm TS}} \nonumber \\
	& = &
	\sigma_{\rm w} \beta_{\rm w} \frac{u_{\rm w}}{u_{\rm TS}}
	\left[
		1 + \frac{\hat{\Gamma}}{4 (\hat{\Gamma} - 1)} \frac{u_{\rm w}}{\gamma_{\rm TS}} \left(1 + \sigma_{\rm w} (1 - \beta^{-1}_{\rm TS}) \right)
	\right]^{-1}, \label{eq:app:RHConditionSigma}
\end{eqnarray}
where the subscripts `w' and `TS' denote the upstream and downstream variables, respectively.
The upstream variables $n_{\rm w},~b_{\rm w},$ and $u_{\rm w}(\gamma_{\rm w})$ are related with $\sigma_{\rm w},~\kappa_{\rm w}$ and $L_{\rm spin}$ as
\begin{eqnarray}
	\sigma_{\rm w}
	& = &
	\frac{(\gamma_{\rm w} b_{\rm w})^2}{u_{\rm w} \gamma_{\rm w} n_{\rm w} m_{\rm e} c^2}
	,\label{eq:app:SigmaWind} \\
	\kappa_{\rm w} \dot{N}_{\rm GJ}
	& = &
	4 \pi r^2_{\rm TS} n_{\rm w} u_{\rm w} c
	,\label{eq:app:NumberFlux} \\
	L_{\rm spin}
	& = &
	\kappa_{\rm w} \dot{N}_{\rm GJ} \gamma_{\rm w} m_{\rm e} c^2 (1 + \sigma_{\rm w}),\label{eq:app:Luminosity}
\end{eqnarray}
where $r_{\rm TS}$ is the radius of the termination shock and $\gamma^2_{\rm w} = 1 + u^2_{\rm w}$.
For notational convenience, the electromagnetic field is divided by $\sqrt{4 \pi}$ and the velocity is normalized by the speed of light $c$.

Following \citet{Komissarov12}, we introduce the fast magnetosonic Mach number at the upstream
\begin{eqnarray}\label{eq:app:MachNumber}
	M_{\rm w}
	& \equiv &
	\frac{u_{\rm w}}{u_{\rm f}},
\end{eqnarray}
where the magnetosonic velocity at the upstream is written as
\begin{eqnarray}\label{eq:app:FastModeVelocity}
	u^2_{\rm f}
	=
	\sigma_{\rm w}.
\end{eqnarray}
$M_{\rm w} \gg 1$ is the condition for a strong shock.
Considering $\sigma_{\rm w} \gg 1$ and $u_{\rm w} \approx \gamma_{\rm w} \gg 1$, the termination shock is strong when
\begin{eqnarray}\label{eq:app:StrongShockCondition}
	\kappa_{\rm w} \sigma^{\frac{3}{2}}_{\rm w}
	& < &
	1.4 \times 10^{10}
	\left( \frac{L_{\rm spin}}{10^{38}~{\rm erg~s^{-1}}} \right)^{\frac{1}{2}},
\end{eqnarray}
where equation (\ref{eq:WindProperties}) is used in order to eliminate $u_{\rm w}$.
We set $\kappa_{\rm w} = 10^4$ and $L_{\rm spin} = 4.6 \times 10^{38}~{\rm erg~s^{-1}}$ in this paper and then we require $\sigma_{\rm w} \ll 2.1 \times 10^4$.
For $1 \ll \sigma_{\rm w} \ll 10^4$ corresponding to the high-$\sigma_{\rm w}$ cases in this paper, we obtain \citep[c.f.,][]{Komissarov12}
\begin{eqnarray}
	u^2_{\rm TS}
	& \approx &
	\sigma_{\rm w}
	, \label{eq:app:HighSigmaBoundaryVelocity} \\
	\frac{n_{\rm TS}}{n_{\rm w}}
	=
	\frac{b_{\rm TS}}{b_{\rm w}}
	& \approx &
	M_{\rm w}
	, \label{eq:app:HighSigmaCompressionRatio} \\
	p_{\rm TS}
	& \approx &
	\frac{1}{8} n_{\rm w} m_{\rm e} c^2 M^2_{\rm w}
	, \label{eq:app:HighSigmaBoundaryPressure} \\
	\sigma_{\rm TS}
	& \approx &
	2 \sigma_{\rm w}. \label{eq:app:HighSigmaBoundarySigma}
\end{eqnarray}
%

\section{Derivation of basic equations}\label{app:Derivation}

\subsection{Ideal MHD equations for a relativistically-hot plasma}\label{app:IdealMHD}

We start from the ideal MHD equations.
The energy-momentum tensor for ideal MHD is composed of a fluid part $T^{\mu \nu}_{\rm FL}$ and an electromagnetic part $T^{\mu \nu}_{\rm EM}$, i.e.,
\begin{eqnarray}\label{eq:app:EnergyMomentumTensor}
	T^{\mu \nu}
	\equiv
	T^{\mu \nu}_{\rm FL} + T^{\mu \nu}_{\rm EM}
	\equiv
	\left(w + b^2 \right) u^{\mu} u^{\nu} + \left( p + b^2 / 2 \right) g^{\mu \nu} - b^{\mu} b^{\nu}.
\end{eqnarray}
We assume the flat space-time.
The magnetic field four-vector $b_{\mu} = (1/2) e_{\mu \nu \alpha \beta} F^{\alpha \beta} u^{\nu}$ ($F_{\alpha \beta} = - e_{\alpha \beta \mu \nu} b^{\mu} u^{\nu}$) is written with the use of the Levi-Civita tensor $e_{\mu \nu \alpha \beta}$, the electromagnetic field $F^{\mu \nu}$ and a four-velocity of the fluid $u^{\mu} = \gamma (1, {\bm \beta})$.
The ideal MHD condition corresponds to the electric field four-vector being zero ($F_{\mu \nu} u^{\nu} = 0$).
The proper enthalpy density $w$ is equal to the sum of the rest mass energy $\rho c^2$, the internal energy $e_{\rm int}$ and the pressure $p$.
Assuming that the fluid satisfies the equation of state,
\begin{eqnarray}\label{eq:app:EOS}
	p
	=
	(\hat{\Gamma} - 1) e_{\rm int},
\end{eqnarray}
with the constant adiabatic index $\hat{\Gamma}$ and composed of electron-positron plasma (the mass of an electron $m_{\rm e}$), we write
\begin{eqnarray}\label{eq:app:EnthalpyDensity}
	w
	=
	n m_{\rm e} c^2 + \frac{\hat{\Gamma}}{\hat{\Gamma} - 1} p,
\end{eqnarray}
where $n$ is the proper number density.

In addition to equation (\ref{eq:app:EnthalpyDensity}), an MHD system is fully described by
\begin{eqnarray}
	\nabla_{\mu} (n u^{\mu})
	& = &
	0
	, \label{eq:app:NumberConservation} \\
	\nabla_{\mu} T^{\mu \nu}
	& = &
	0
	, \label{eq:app:EnergyMomentumConservation} \\
	e^{\mu \nu \alpha \beta} \nabla_{\nu} F_{\alpha \beta}
	& = &
	0
	, \label{eq:app:InductionEquation}
\end{eqnarray}
where the conservation of the particle number, the conservations of the energy and momentum, and the induction equation, respectively.

\subsection{Radial outflow with pure toroidal magnetic field: KC model}\label{app:KC84}

\citet{Kennel&Coroniti84a} considered the radial outflow $u^{\mu} = (\gamma, u {\bm e}_r)$ with the (ordered) toroidal magnetic field $\bar{b}^{\mu} = (0, \bar{b} {\bm e}_{\phi})$.
Now, with the use of Equation (\ref{eq:app:EnthalpyDensity}), we have only four variables of the system, the radial velocity $u = \gamma \beta$, the toroidal field strength $\bar{b} = \bar{B} / \gamma$, the number density $n$, and the enthalpy density $w$ (relativistically hot plasma $\hat{\Gamma} = 4/3$), where $\bar{B}$ is the magnetic field in the observer frame.
For corresponding four equations, they chose
\begin{eqnarray}
	\nabla_{\mu} (n u^{\mu})
	& = &
	0
	, \label{eq:app:NumberConservationKC84} \\
	\nabla_{\mu} T^{\mu t}
	& = &
	0
	, \label{eq:app:TotalEnergyConservationKC84} \\
	u_{\nu} \nabla_{\mu} T^{\mu \nu}_{\rm FL}
	& = &
	0
	, \label{eq:app:InternalEnergyConservationKC84} \\
	e^{\phi \nu \alpha \beta} \nabla_{\nu} F_{\alpha \beta}
	& = &
	0
	. \label{eq:app:InductionEquationKC84}
\end{eqnarray}
In the ideal MHD approximation, the contraction $u_{\nu} \nabla_{\mu} T^{\mu \nu} = 0$ is satisfied separately for the fluid (equation (\ref{eq:app:InternalEnergyConservationKC84})) and for the electromagnetic parts, and these equations describe the propagation of the internal energy of the fluid (the first law of the thermodynamics) and of the magnetic energy density, respectively.
Actually, the contraction of the induction equation with the magnetic field is equivalent to the propagation of the magnetic energy density, i.e.,
\begin{eqnarray}
	\frac{1}{2} b_{\mu} e^{\mu \nu \alpha \beta} \nabla_{\nu} F_{\alpha \beta}
	=
	- u_{\nu} \nabla_{\mu} T^{\mu \nu}_{\rm EM}
	& = &
	0
	. \label{eq:app:MagneticEnergyConservationKC84}
\end{eqnarray}
%

\subsection{Adding radiative loss and phenomenological magnetic dissipation terms to KC model}\label{app:DS02}

\citet{Drenkhahn&Spruit02} extended the formulation by including the radiation loss and the magnetic dissipation terms.
The radiation extracts the energy from the fluid part.
Here, we introduce the comoving emissivity (radiation loss rate) $\Lambda_{\rm rad}$.
The energy and momentum are no more conserved and then equation (\ref{eq:app:EnergyMomentumConservation}) becomes
\begin{eqnarray}
	\nabla_{\mu} T^{\mu \nu}
	& = &
	- u^{\nu} \Lambda_{\rm rad} / c
	. \label{eq:app:RadiativeLoss}
\end{eqnarray}

\citet{Drenkhahn02} introduced dissipation of the magnetic energy into the ideal MHD equations (equations (\ref{eq:app:NumberConservationKC84}) -- (\ref{eq:app:InductionEquationKC84})).
Note that they did not deduce the magnetic dissipation term from the resistive MHD formulation, but adopted a rather phenomenological and straightforward way.
We adopt their phenomenological formulation of magnetic dissipation in this paper.
We only have to add a magnetic dissipation term $- (b^2 / 2) / \tau_{\rm diss}$ to the right-hand side of equation (\ref{eq:app:MagneticEnergyConservationKC84}) and the corresponding heating term is added to equation (\ref{eq:app:InternalEnergyConservationKC84}) in order to preserve equation (\ref{eq:app:RadiativeLoss}), where $\tau_{\rm diss}$ is the dissipation time-scale in the proper frame.
Finally, the four equations describing such a system are
\begin{eqnarray}
	\nabla_{\mu} (n u^{\mu})
	& = &
	0
	, \label{eq:app:NumberConservationDS02} \\
	\nabla_{\mu} T^{\mu t}
	& = &
	- \gamma \Lambda_{\rm rad} / c
	, \label{eq:app:EnergyEquationDS02} \\
	- u_{\nu} \nabla_{\mu} T^{\mu \nu}_{\rm FL}
	& = &
	\frac{b^2 / 2}{c \tau_{\rm diss}}
	- \frac{\Lambda_{\rm rad}}{c}
	, \label{eq:app:InternalEnergyEquationDS02} \\
	- u_{\nu} \nabla_{\mu} T^{\mu \nu}_{\rm EM}
	& = &
	- \frac{b^2 / 2}{c \tau_{\rm diss}}
	. \label{eq:app:MagneticEnergyEquationDS02}
\end{eqnarray}
%

\subsection{Turbulent magnetic field}\label{app:TurbulentMagneticField}

Here, we describe our treatment of the turbulent magnetic field.
The turbulent magnetic field is set to be isotropic in the proper frame and then the magnetic field three-vector in the proper frame would be written as ${\bm b} = \bar{b} {\bm e}_{\phi} + \delta {\bm b}$ satisfying $\langle {\bm b} \rangle = \bar{b} {\bm e}_{\phi}$ and $\langle {\bm b}^2 \rangle = b^2_{\phi} + \delta b^2$, where $\langle \rangle$ represents ensemble average.
Each component of the turbulent magnetic field $\delta {\bm b} = \delta b_r {\bm e}_r + \delta b_{\theta} {\bm e}_{\theta} + \delta b_{\phi} {\bm e}_{\phi}$ has the same amplitude and
\begin{eqnarray}\label{eq:app:TurbulentMagneticField}
	\langle \delta b_i \delta b_j \rangle
    & = &
    \frac{\delta b^2}{3} \delta_{i j},
\end{eqnarray}
where $i,j = r, \theta, \phi$ and $\delta_{i j}$ is the Kronecker delta.

In the observer (pulsar) frame, the flow is radial (${\bm u} = u {\bm e}_r$) and thus the magnetic field in this frame is
\begin{eqnarray}\label{eq:app:TotalMagneticThreeVector}
	{\bm B}
    & = &
    {\bm b}_{\parallel} + \gamma {\bm b}_{\perp}, \nonumber \\
    & = &
    \delta b_{r} {\bm e}_{r} + \gamma \delta b_{\theta} {\bm e}_{\theta} + \gamma (\bar{b} + \delta b_{\phi}) {\bm e}_{\phi},
\end{eqnarray}
where only the components which are perpendicular to the flow velocity are amplified by the Lorentz contraction.
Finally, we find the magnetic four-vector including the turbulent magnetic field, which is written as
\begin{eqnarray}\label{eq:app:TotalMagneticFourVector}
	b^{\mu}
	& = &
    (({\bm u} \cdot {\bm B}), ({\bm B} + ({\bm u} \cdot {\bm B}) {\bm u}) / \gamma), \nonumber \\
    & = &
	(u \delta b_r, \gamma \delta b_r, \delta b_{\theta}, \bar{b} + \delta b_{\phi}),
\end{eqnarray}
where the ideal MHD approximation is used in order to obtain the first line.
It is easy to identify the ordered ($\bar{b}^{\mu} = \bar{b} {\bm e}_{\phi}$) and turbulent ($\delta b^{\mu} \equiv b^{\mu} - \bar{b}^{\mu}$) magnetic field from equation (\ref{eq:app:TotalMagneticFourVector}) and they satisfy
\begin{eqnarray}\label{eq:app:UsefulRelations}
	\bar{b}^{\mu} u_{\mu} = 0,~\delta b^{\mu} u_{\mu} = 0,
\end{eqnarray}
separately.
$\langle b^2 \rangle = \langle b^{\mu} b_{\mu} \rangle = \bar{b}^2 + \delta b^2$ is also immediately found from equations (\ref{eq:app:TurbulentMagneticField}) and (\ref{eq:app:TotalMagneticFourVector}).

\subsection{Our formulation including turbulent magnetic field}\label{app:OurFormulation}

We further add the turbulent magnetic field $\delta b^{\mu}$ as a simple extension of the equations given by \citet{Drenkhahn&Spruit02} (equations (\ref{eq:app:NumberConservationDS02}) $-$ (\ref{eq:app:MagneticEnergyEquationDS02})).
We consider dissipation of the turbulent component of magnetic field in the nebular flow rather than the annihilation of the anti-parallel toroidal components of the striped wind.
The turbulent components would be converted from the toroidal ones.
We need another equation to determine evolution of the additional variable $\delta b^2$

The equations which describe evolution of $\bar{b}^2$ and $\delta b^2$ are found from the contractions of the induction equation (equation (\ref{eq:app:InductionEquation})) with $\bar{b}_{\mu}$ and $\delta b_{\mu}$, respectively.
The equations would be written as
\begin{eqnarray}
	\langle \nabla_{\mu} (n u^{\mu}) \rangle
	& = &
	0
	, \label{eq:app:NumberConservationTTT} \\
	\langle \nabla_{\mu} T^{\mu t} \rangle
	& = &
	- \gamma \Lambda_{\rm rad} / c
	, \label{eq:app:TotalEnergyConservationTTT} \\
	- \langle u_{\nu} \nabla_{\mu} T^{\mu \nu}_{\rm FL} \rangle
	& = &
	\frac{\delta b^2 / 2}{c \tau_{\rm diss}}
	- \frac{\Lambda_{\rm rad}}{c}
	, \label{eq:app:InternalEnergyConservationTTT} \\
	\frac{1}{2} \langle \bar{b}_{\mu} e^{\mu \nu \alpha \beta} \nabla_{\nu} F_{\alpha \beta} \rangle
	& = &
	- \frac{\bar{b}^2 / 2}{c \tau_{\rm conv}}
	, \label{eq:app:ToroidalMagneticEnergyConservationTTT} \\
	\frac{1}{2} \langle \delta b_{\mu} e^{\mu \nu \alpha \beta} \nabla_{\nu} F_{\alpha \beta} \rangle
	& = &
	- \frac{\delta b^2 / 2}{c \tau_{\rm diss}}
	+ \frac{\bar{b}^2  / 2}{c \tau_{\rm conv}}
	. \label{eq:app:TurbulentMagneticEnergyConservationTTT}
\end{eqnarray}
The sum of equations (\ref{eq:app:InternalEnergyConservationTTT}) $-$ (\ref{eq:app:TurbulentMagneticEnergyConservationTTT}) corresponds to $- \langle u_{\nu} \nabla_{\mu} T^{\nu \mu} \rangle = - \Lambda_{\rm rad} / c$ (see equation (\ref{eq:app:RadiativeLoss})).
After the straightforward calculation, we obtain equations (\ref{eq:ToroidalMagneticEnergyConservation}) and (\ref{eq:TurbulentMagneticEnergyConservation}) from equations (\ref{eq:app:ToroidalMagneticEnergyConservationTTT}) and (\ref{eq:app:TurbulentMagneticEnergyConservationTTT}), respectively.
Although we apply these equations to the flow deceleration in this paper, they will also be applicable to the flow acceleration \citep[c.f.,][]{Heinz&Begelman00}.


\bsp	
\label{lastpage}
\end{document}